\documentclass[prx,twocolumn,aps,epsf,showpacs,superscriptaddress,longbibliography,footinbib]{revtex4-1}
\usepackage[pdftex]{graphicx}

\usepackage{dcolumn}
\usepackage{bm}
\usepackage{epsfig}
\usepackage{latexsym} 
\usepackage{amsmath}
\usepackage{amssymb}
\usepackage{color}
\usepackage{array}
\usepackage{bbm}
\usepackage{hyperref}
\usepackage{color}
\usepackage{array}
\usepackage{xcolor}

\usepackage{booktabs}
\newcolumntype{C}[1]{>{\centering\arraybackslash}m{#1}}
\AtBeginDocument{
\heavyrulewidth=.08em
\lightrulewidth=.05em
\cmidrulewidth=.03em
\belowrulesep=.65ex
\belowbottomsep=0pt
\aboverulesep=.4ex
\abovetopsep=0pt
\cmidrulesep=\doublerulesep
\cmidrulekern=.5em
\defaultaddspace=.5em
}
\usepackage{booktabs}
\newcolumntype{R}[1]{>{\raggedleft\arraybackslash}p{#1}}
\AtBeginDocument{
\heavyrulewidth=.08em
\lightrulewidth=.05em
\cmidrulewidth=.03em
\belowrulesep=.65ex
\belowbottomsep=0pt
\aboverulesep=.4ex
\abovetopsep=0pt
\cmidrulesep=\doublerulesep
\cmidrulekern=.5em
\defaultaddspace=.5em
}

\newcommand{\<}{\langle}
\newcommand{\e}{\varepsilon}

\renewcommand{\>}{\rangle}
\renewcommand{\(}{\left(}
\renewcommand{\)}{\right)}
\renewcommand{\[}{\left[}
\renewcommand{\]}{\right]}
\renewcommand{\v}[1]{\boldsymbol{#1}} 

\renewcommand{\d}{\partial}

\newcommand{\eps}{\epsilon}

\begin{document}
\title{Fractal non-Fermi liquids from moir\'e-Hofstadter phonons}
\author{Ajesh Kumar}
\email{These authors contributed equally to this work}
\affiliation{Department of Physics, University of Texas at Austin, Austin, TX 78712, USA}

\author{Zihan Cheng}
\email{These authors contributed equally to this work}
\affiliation{Department of Physics, University of Texas at Austin, Austin, TX 78712, USA}

\author{Andrew C. Potter}
\affiliation{Department of Physics, University of Texas at Austin, Austin, TX 78712, USA}

\begin{abstract}
We theoretically explore 2d moir\'e heterostructures in lattice-commensurate magnetic fields as platforms for quantum simulation of a paradigmatic model of non-Fermi liquid physics: a Fermi-surface coupled to a fluctuating gauge field. In these moir\'e-Hofstadter (MH) systems, long-wavelength acoustic phonons exhibit singular interactions with electrons analogous to those of electrons with 2d gauge fields. This leads to a breakdown of Fermi-liquid theory at low temperatures. We show that a combination of large moir\'e-unit cell size, tunable Fermi-surface topology, and enhanced coupling to interlayer sliding modes, enhance these effects by over many orders-of-magnitude compared to bulk crystals, placing them within experimental reach. Though we find that the asymptotic low-temperature non-Fermi liquid regime remains at prohibitively low temperatures, striking precursor non-Fermi liquid signatures can be observed, and we propose surface acoustic wave attenuation and quantum oscillation transport experiments. We also study the motion of MH acoustic (MHA)-polarons, which we predict exhibit logarithmically diverging effective mass and unconventional magnetic field scaling for scaling of cyclotron resonance frequency and quantum oscillation amplitude.
\end{abstract}
\maketitle

\section{Introduction}
The problem of metallic electrons strongly coupled to fluctuating gapless (bosonic) collective modes is believed to underly some of the least-understood quantum phenomena, from strange-metal phases of high-temperature superconductors~\cite{Lee1989,lee1992gauge,Lee2006}, to metallic quantum critical systems~\cite{Metlitski2010,Metlitski2010a}, composite fermion liquids~\cite{Halperin1993}, and gapless spin-liquids~\cite{Lee2005,Senthil2008,Lee2008,Lee2009}. These systems lack well-defined quasi-particles and are not captured by conventional Fermi liquid theory paradigm. The detailed behavior of these non-Fermi liquid (NFL) systems remain poorly understood due to the absence of naturally controlled theoretical calculations~\cite{Lee2009,Mross2010,Dalidovich2013} or efficient numerical methods, complex materials chemistry, relatively high impurity concentrations, and limited ability to tune the electron density or interactions in the underlying host materials. To this end alternative platforms to explore NFL behavior in simpler, cleaner, and more tunable material platforms are highly desirable. 

A common and reliable source of gapless bosonic collective modes are: Nambu-Goldstone modes (NGM), such as acoustic phonons and magnons, arising due to spontaneously broken continuous symmetries. However, the same mechanism that ensures their masslessness ordinarily causes NGM to decouple from electrons at low temperatures, producing conventional Fermi-liquid behavior. Exceptions occur for rotational NGM, including nematic NGM~\cite{Oganesyan2001} and magnons in spin-orbit coupled metals~\cite{Xu2010,Bahri2015}, for which the electron-NGM coupling does not freeze out at low temperatures. However, these examples require continuous rotation symmetry, a situation that can be at best approximately realized in crystalline materials. 

Watanabe and Vishwanath~\cite{Watanabe2014} derived a general criterion for exceptional NGM, and pointed out another, rather surprising, example: phonons of a crystal in a magnetic field. The non-commutative nature of translations in a magnetic field result in electrons coupling to phonon fluctuations in much the same way as they would couple to dynamically fluctuating magnetic fields. An important caveat is that magnetic fields tend to produce non-dispersing Landau levels, destroying the Fermi-surface at the single-particle level. In a crystal, the bandwidth of Landau levels, which sets an upper bound on the energy scale for observing NFL behavior, scales as $e^{-1/\nu}$ where $\nu$ is the number of flux per unit cell. For atomic-scale lattices, reaching $\nu\sim 1$, would require astronomical $B\gtrsim 10^4$T. For this reason, the prediction of~\cite{Watanabe2014} has not been experimentally tested.

In contrast, moir\'e-superlattice potentials of small-angle twisted structures can have sufficiently large unit-cells to reach $\nu\sim \mathcal{O}(1)$ with laboratory magnetic fields. In this paper we explore these systems as platforms for exploring NFL physics, focusing on twisted bilayer graphene (TBG) as a particularly promising example due to its large moir\'e-potential.

\begin{figure*}[t]
	\includegraphics[width=\textwidth]{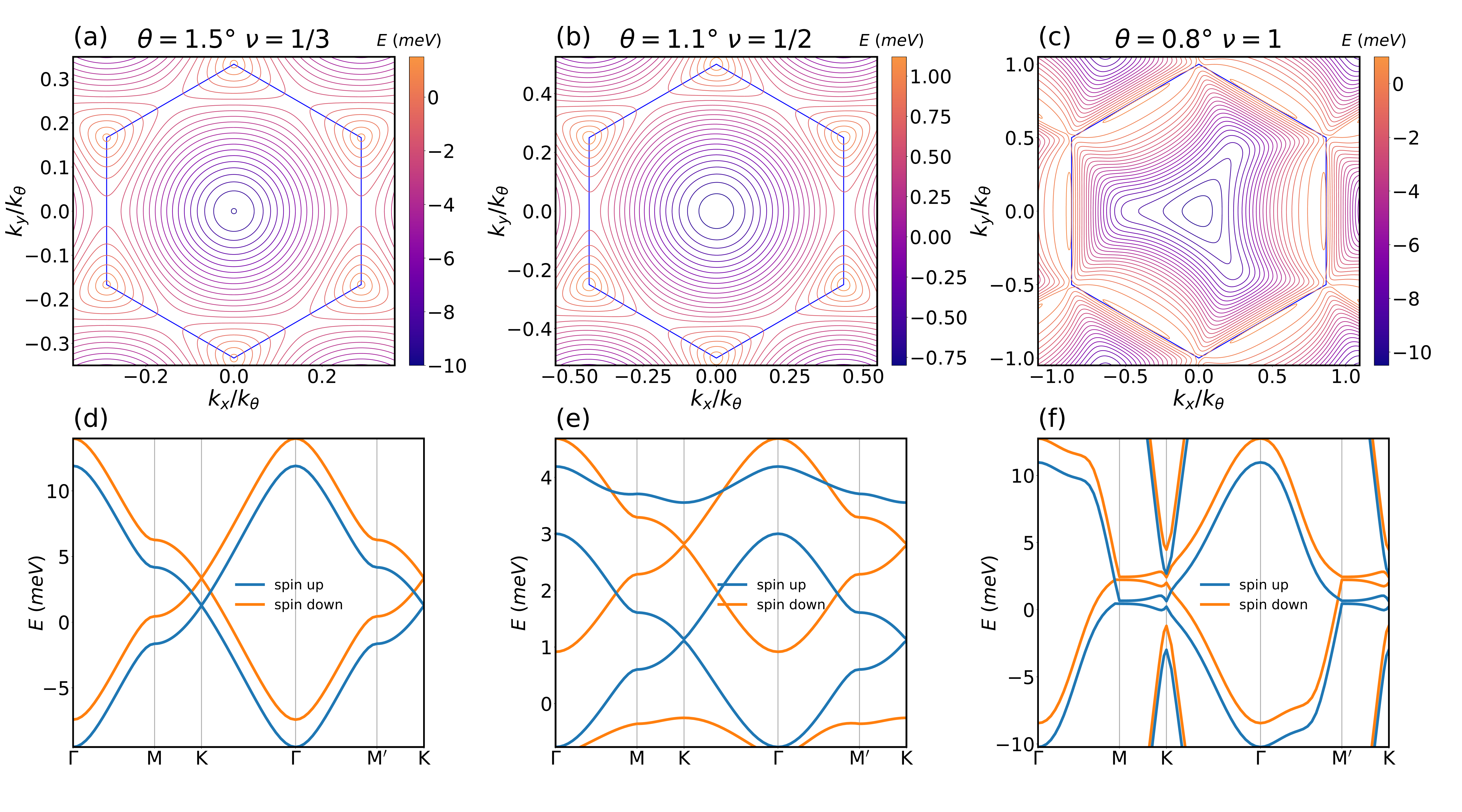}
	\caption{{\bf Magnetic band structure -- } (a), (b), (c) Magnetic bands structures of spin up valence bands in the magnetic Brillouin zone for $\theta=1.5^\circ$, $\nu=1/3$, $B=18.0T$, $\theta=1.1^\circ$, $\nu=1/2$, $B=14.5T$ and $\theta=0.8^\circ$, $\nu=1$, $B=15.4T$. In (a) and (b) there is a pair of Dirac points at $K, K'$ and an electron pocket at $\Gamma$ point, whereas in (c) Dirac points at $K, K'$ smear out to oblong hole pockets near $M, M'$, forming a three-fold Fermi surface. The magnetic Brillouin zone is shown as a blue hexagon. (d), (e), (f) bands structure along the high symmetry path $\Gamma-{\rm M}-{\rm K}-\Gamma-{\rm M^{\prime}}-{\rm K}$ for $\theta=1.5^\circ$, $\nu=1/3$, $\theta=1.1^\circ$, $\nu=1/2$ and $\theta=0.8^\circ$, $\nu=1$.}
	\label{fig:bands} 
	\vspace{-0.2in}
\end{figure*}

\section{Moir\'e-Hofstadter (MH) bands and phonons}
\subsection{Electronic band structure}
Our analysis begins from Bistritzer and Macdonald's (BM) continuum model for twisted bilayer graphene (TBG)~\cite{Bistritzer2011,Bistritzer2011a,Hejazi2019}, with single-electron Hamiltonian:
\begin{align}
    H(\boldsymbol{r})=\begin{pmatrix}
    h(-\theta/2)& T(\boldsymbol{r})\\
    T^\dagger(\boldsymbol{r})& h(\theta/2)
    \end{pmatrix},
    \label{eq:rbmodel}
\end{align}
where $h(\theta)= v\(\boldsymbol{\Pi}+{\rm sgn}(\theta)\frac{\boldsymbol{k}_\theta}{2}\)\cdot \boldsymbol{\sigma}_\theta$ is the Hamiltonian for a single graphene layer twisted by angle $\theta$, $\boldsymbol{\Pi}=\boldsymbol{p}+e\boldsymbol{A}$ is the canonical momentum with in magnetic field $\v{B} = \nabla\times \v{A} = B\hat{\v{z}}$, $\v{k}_\theta= -k_\theta\hat{\v{y}}\equiv -\frac{8\pi}{3a}\sin\frac{\theta}{2}\hat{\v{y}}$ is the vector connecting the Dirac points in the two layers, and $a$ is the graphene lattice spacing. $T(\v{r})$ represent interlayer tunneling with the (approximate) spatial periodicity of the moir\'e lattice
\begin{align}
	T(\boldsymbol{r})=w\sum_je^{-i\boldsymbol{g}_j\cdot\boldsymbol{r}}T_j.
\end{align}
Here $\boldsymbol{g}_0=0$, $\boldsymbol{g}_{1,2}=\frac{\sqrt{3}k_\theta}{2}\(\mp\hat{x}+\sqrt{3}\hat{y}\)$ are moir\'e-reciprocal-lattice vectors, and $T_j = \eta+ \sigma_{2\pi j/3}$ are sub-lattice matrices. The real-space lattice vectors for the moir\'e cell are (for small twist angle, $\theta\ll 1$): $\v{a}_{1,2}= \frac{a}{2\theta}\(\mp\sqrt{3}\hat{x}+\hat{y}\)$.
For numerical results presented below, we take $\hbar v\approx 610\ {\rm meV\ nm}$, $w\approx 110\ {\rm meV}$, and $\eta\approx 0.82$, which accounts corrugation~\cite{Koshino2018}. 

Varying $B$ produces a fractal ``Hofstadter butterfly"
energy spectrum~\cite{Bistritzer2011a,Hejazi2019,Dean2013,Hunt2013}. Here, instead, we  focus on fixed $B$ 
such that number of magnetic flux per moir\'e unit cell is a rational fraction~\cite{Bistritzer2011a}\ (see also Appendix \ref{app:mhbands}): 
\begin{align}
\nu = \frac{\sqrt{3}a^2}{4\pi \theta^2 \ell_B^2}=\frac{p}{q}, ~~\text{with}~~ p,q\in \mathbb{Z},
\end{align}
and magnetic length $\ell_B = 1/\sqrt{eB}$. At these commensurate fields, the Hamiltonian possesses a periodic magnetic-lattice translation symmetry generated by $\{\v{a}_1,q\v{a}_2\}$, and exhibits dispersive Bloch-bands labeled by magnetic quasi-momenta.

\begin{figure*}[t]
	\includegraphics[width=\textwidth]{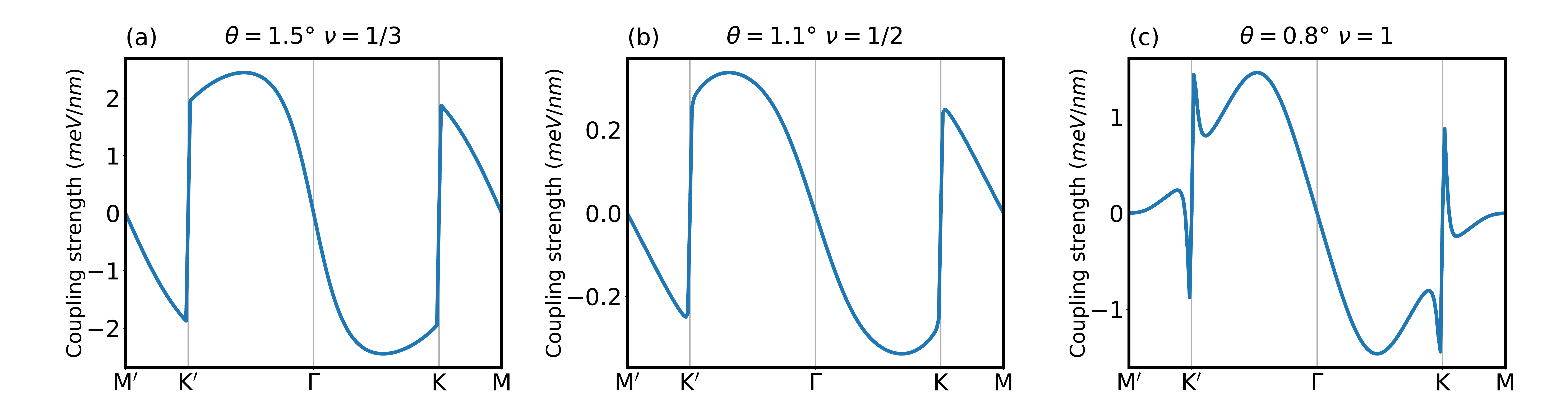}
	\caption{{\bf Electron-phonon coupling -- } electron-phonon coupling $\Gamma_{\rm m}$, for $x$-polarization phonons in the limit of zero momentum transfer for (a) $\theta=1.5^\circ$, $\nu=1/3$, (b) $\theta=1.1^\circ$, $\nu=1/2$ and (c) $\theta=0.8^\circ$, $\nu=1$.}
	\label{fig:coupling} 
	\vspace{-0.2in}
\end{figure*}

Fig.~\ref{fig:bands}(a) and (d) show the lowest two bands near charge-neutrality for $\theta=1.5^\circ$, and $B=18T$ ($\nu=\frac13$). Apart from having lower bandwidth and spin-splitting, the bands resemble those of single-layer graphene in zero-field. 
In particular, the conduction (valence) bands exhibits a pair of Dirac points near charge-neutrality at the moir\'e $K,K'$ points, which merge into a single $\Gamma$-centered hole (electron) pocket at the top of the conduction band (bottom of the valence band), crossing through a van-Hove (VH) singularity with nearly-nested hexagonal Fermi-surface at intermediate filling along the way. Bands at $\theta=1.1^\circ$ (shown in Fig.~\ref{fig:bands}(b) and (e)) exhibit a similar Fermi surface evolution with doping, and also inherit the zero-magnetic field property of having reduced bandwidth near the magic angle~\cite{Bistritzer2011}. In this case, because of strong electron-electron interactions relative to the small bandwidth, correlated insulating and superconducting behaviors have been observed~\cite{cao2018unconventional}. However, at the magnetic fields that we consider here, the superconductivity near the magic angle is suppressed~\cite{cao2018unconventional}, and the correlated insulating behavior arises only near integer fillings of the flat moir\'e bands~\cite{cao2018correlated,lu2019superconductors}, which can be generically avoided~\footnote{The non-Fermi liquid effects that we find are enhanced near the VH fillings, which can be tuned to generic non-integer fillings via an in-plane (Zeeman) field.}. 
Therefore, we expect the correlated insulating and superconducting behaviors do not necessarily obscure the non-Fermi liquid physics discussed in this work.

At special flat Fermi-surface (FFS) fillings, which occur near (but not precisely at) VH fillings, the Fermi-surface curvature vanishes at certain points on the Fermi-surface, and remains small over extended patches, which will enhance electron-phonon interaction effects. Taylor expanding the dispersion in small momentum displacements from the center of these flat-patches, the vanishing Fermi-surface curvature requires a vanishing quadratic term for the tangential dispersion. Due to mirror symmetry around the center of the flat-patch, the next leading term is quartic. Near the flat-patches both the quadratic and the quartic terms contribute to the dispersion:
\begin{align}
\epsilon_{\v{k}} = v_F k_{\perp} + \frac{1}{2m} k_{\parallel}^2 + \lambda k_{\parallel}^4
\end{align}
where $k_\perp$ and $k_{\parallel}$ are patch-momentum components perpendicular and parallel to the Fermi-surface. Away from the FFS points, there is a crossover momentum scale $q_* = \frac{1}{\sqrt{2m \lambda}}$, below which the leading dependence on $k_{\parallel}$ is $\epsilon_{\v{k}} \sim k_{\parallel}^2$ and above which it is $\epsilon_{\v{k}} \sim k_{\parallel}^4$.

In SLG, FFS fillings were predicted to enable correlated insulators or superconductors~\cite{Nandkishore2012}, but occur at prohibitively high electron density $\sim 10^{15}{\rm cm}^{-2}$  above charge neutrality. In moir\'e systems, the FFS occur at much lower fillings and can be experimentally explored. Moreover, the degree of Ferm-surface nesting (alignment between antipodal flat-patches), which controls the competition between interaction-driven orders, can be adjusted by twist-engineering. For example, for the parameters shown in Fig.~\ref{fig:bands}(c), the FFS fillings exhibit a non-nested FFS with three-fold rotational symmetry.

\subsection{Electron-phonon coupling}
To describe phonons, we introduce displacement fields: $\v{u}_\ell(\v{r})$ for each layer $\ell=1,2$, which are conveniently re-expressed in terms of the mean-displacement of the bilayer: $\bar{\v{u}} = \frac12\(\v{u}_1+\v{u}_2\)$ and relative displacements of the layers, $\v{d} = \v{u}_1-\v{u}_2$. Both types of phonon displacements on the electronic Hamiltonian can be accounted by displacing the tunneling operators $T(\v{r})\rightarrow T(\v{r}-\v{u})$ by 
with a single effective displacement field~\cite{Bistritzer2011,Balents2019,Lian2019}:
\begin{align}
	\v{u} = \bar{\v{u}}-\frac{\hat{z}\times \v{d}}{2\tan\theta/2},
\end{align}

since the relative sliding of the two graphene sheets is equivalent to a translation of the moir\'e pattern perpendicular to the sliding direction. This has two important consequences: First, at small twist angles, the coupling to $\v{d}$ is enhanced by a factor of $\approx \theta^{-1}$~\cite{Lian2019}. Second, for a commensurate crystal relative displacements would be gapped optical modes. Instead, for incommensurate twist angles uniform $\v{d}$-displacements have no energy cost~\cite{koshino2019moire,Wu2019} and result in gapless acoustic modes (in analogy to the sliding ``phason" mode of incommensurate charge density wave orders~\cite{Overhauser1971}). 

The electron-phonon coupling is conveniently identified by changing coordinates to a co-moving frame of the lattice: $\v{r}\leftarrow \(\v{r}-\v{u}(\v{r})\)$~\cite{Balents2019}, in which the lattice is restored to its undistorted form. 
To properly account for the non-commutivity of magnetic translations, while manifestly preserving gauge invariance, we implement this frame transformation with the unitary operator:
\begin{equation}
	W_{\boldsymbol{u}}=e^{-i\boldsymbol{u}(\boldsymbol{R})\cdot\boldsymbol{\Pi}}
\label{eq:Wu}
\end{equation}
where $\v{R} = \v{r}-\ell_B^2\hat{\v{z}}\times \v{\Pi}$ is the guiding center coordinate (whose components all commute with $\v{\Pi}$). Neglecting subleading terms of order $\mathcal{O}(\nabla u, u^2)$, this transformation effectively restores the moir\'e tunneling potential:  $W^\dagger_{\v{u}}T\(\v{r}-\v{u}(\v{r})\)W^{\vphantom\dagger}_{\v{u}} \approx T(\vec{r})$, while transforming the kinetic energy by: $W^\dagger_{\v{u}}\v{\Pi}W^{\vphantom\dagger}_{\v{u}} \approx \v{\Pi}-e\v{B}\times\v{u}(\v{R})$, yielding a direct (gradient-free) electron-phonon interaction:
\begin{equation}
	{H}_\text{e-ph} =  ev\vec{u}\cdot \vec{B}\times
		\begin{pmatrix}
			\hat{\v{\sigma}}_{-\theta/2}& 0\\
			0&\v{\sigma}_{\theta/2}
		\end{pmatrix}
	\equiv \bar{\v{u}}\cdot\v{\Gamma}_{\rm m}+\v{d}\cdot\v{\Gamma}_{\rm r}
\end{equation}
For future convenience, we define the electron phonon vertices:
\begin{align}
\Gamma_{n,\alpha,\lambda}(\v{k},\v{q}) = \<n,\v{k}+\v{q}|\hat{\v{\e}}_\lambda(\v{q}) \cdot\v{\Gamma}_\alpha|n,\v{k}\>
\end{align}
for the $n^\text{th}$ magneto-Bloch band, with wave-vector $\v{k}$ state $|n,\v{k}\>$, where $\alpha\in \{m,r\}$ labels the phonon type, and $\lambda$ labels longitudinal (L) or transverse (T) polarizations corresponding to polarization vector $\hat{\v{\e}}_\lambda(\v{q})$. In the following, we focus on the relative (interlayer-sliding) phonon mode since their coupling to electrons is stronger by a factor $\approx 1/\theta$ compared to the layer-symmetric phonon mode. For notational simplicity we omit the band-index $n$ and the phonon type index $\alpha$ in subsequent expressions. Crucially, these vertices generically do not vanish in the limit of zero momentum transfer~\cite{Watanabe2014} ($\v{q}\rightarrow 0$), as demonstrated numerically in Fig. \ref{fig:coupling}. 

At asymptotically low-temperatures, this direct-coupling is expected to produce a complex non-Fermi liquid state~\cite{Oganesyan2001,Xu2010,Bahri2015,Watanabe2014}, 
The NFL properties cannot be reliably calculated except in artificial limits~\cite{Mross2010,Dalidovich2013}. We will make predictions based only on leading order perturbative calculations, which reliably predict the onset of the non-Fermi liquid behavior approached from higher-temperatures or energy-scales, but are merely suggestive of the possible asymptotic behavior at low-temperatures.

\begin{figure}[t]
	\begin{centering}
	\includegraphics[width=0.47\textwidth]{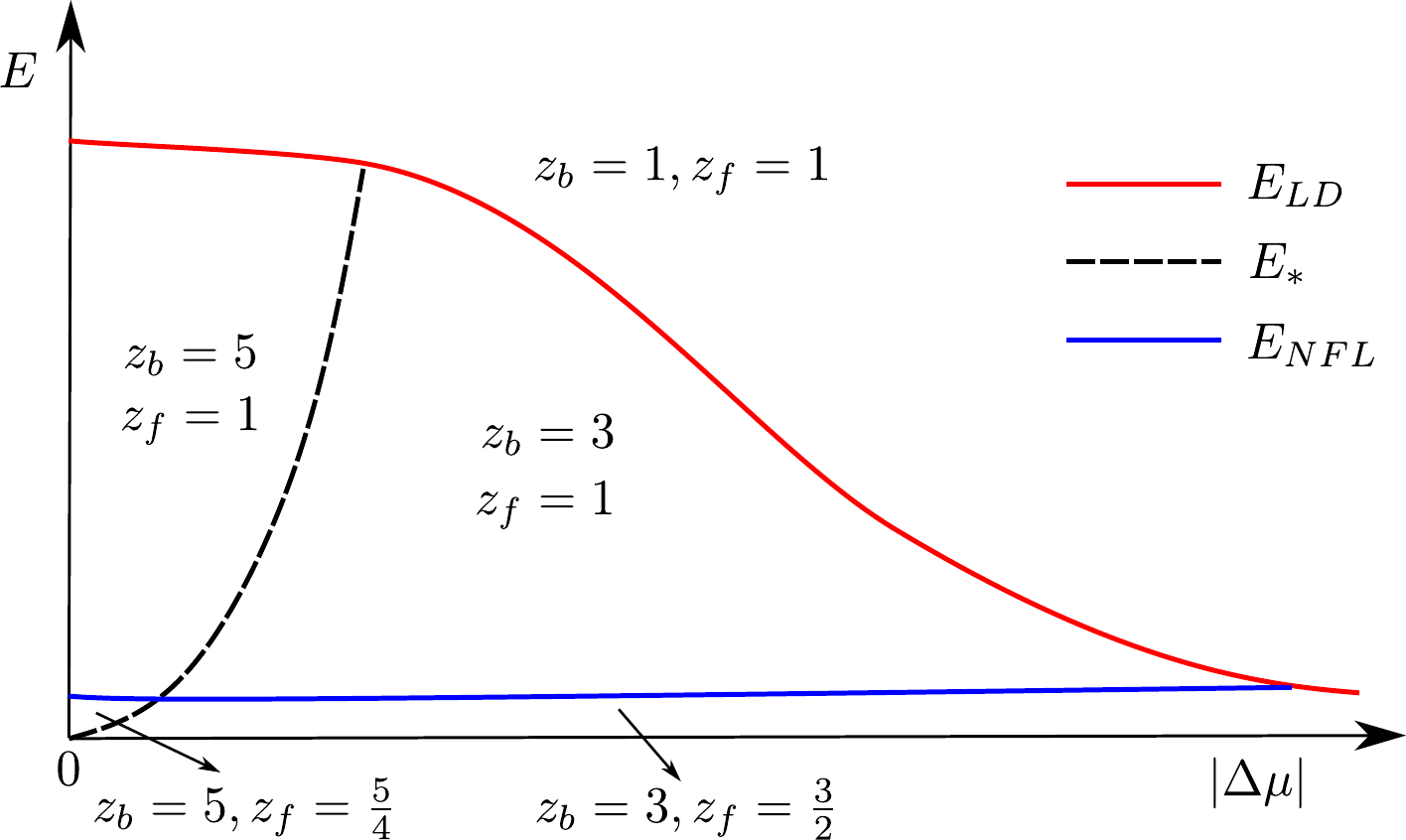}
	\caption{Schematic showing crossover scales $E_\text{LD}$, $E_*$ and $E_{\rm NFL}$. $\Delta \mu$ indicates the deviation in the chemical potential from FFS fillings.}
	\label{fig:schematic} 
	\vspace{-0.2in}
	\end{centering}
\end{figure}

\section{Fractal non-Fermi liquid}
\subsection{Landau damping}
In the presence of a Fermi-surface, this direct coupling causes phonons to decay into the electron-hole continuum, and become soft and overdamped. For simplicity we model the phonon dispersion as isotropic with speed $c_s$ ($\sim 10^4 m/s$)~\cite{cocemasov2013phonons,koshino2019moire}. The effective action in a patch description is:
\begin{align}
S&=\int\frac{d\omega d^2k}{(2\pi)^3}\psi^{ \dagger}_{\omega,\v{k}}(i\omega- \epsilon_{\v{k}})\psi_{\omega,\v{k}}\nonumber\\
&+\sum_{\lambda}\int\frac{d\Omega d^2q}{(2\pi)^3}\rho \(\Omega^2 + c_s^2 q^2 \) \vert u_{\lambda}(\Omega,\v{q})\vert^2\nonumber\\
&+\sum_{\lambda} \int\frac{d\omega d^2kd\Omega d^2q}{(2\pi)^6}\Gamma_{\lambda}(\v{k},\v{q})u_\lambda(\Omega,\v{q})\psi^{\dagger}_{\omega+\Omega,\v{k}+\v{q}}\psi_{\omega,\v{k}}
\end{align}
where $\psi$, $u$ are electron, phonon operators, and $\rho$ is the mass density for SLG. For the dispersion, we use $\epsilon_{\v{k}} \approx v_F k_{\perp} + \frac{1}{2m} k_{\parallel}^2$ away from the FFS patches and $\epsilon_{\v{k}} \approx v_F k_{\perp} + \lambda k_{\parallel}^4$ near the FFS patches. The dynamical critical exponents for the phonons ($z_b$) and the electrons ($z_f$) that we find below are also patch-dependent. In the following, we calculate these exponents individually near and away from the FFS patches, and further provide energy scales for crossovers between these regimes. A self-consistent one-loop perturbative calculation gives a singular Landau-damping (LD) form of the phonon self-energy:
\begin{align}
\Pi_\text{1-loop}(\Omega,\v{q};z_b) =
\begin{cases}
		\gamma_{\lambda}\frac{\vert\Omega\vert}{\vert q_{\parallel}\vert}, & z_b=3 \\
		\gamma_{\lambda} q_*^2 \frac{\vert\Omega\vert}{\vert q^3_{\parallel}\vert}, & z_b=5
	\end{cases}
\end{align}
where $q_{\parallel}$ is the component of $q$ parallel to the Fermi-surface,  and we define the LD-parameters:
\begin{equation}
	\gamma_{\lambda}=\frac{\vert m\vert}{\pi \rho v_F}\left.\Gamma^2_{\lambda}(\v{k},\hat{{q}})\right\vert_{\hat{\v{q}}\parallel {\rm FS}}
\end{equation}
The expressions are labeled by the resulting dynamical critical exponent $z_b=3,5$ describing the low wave-vector dependence of the phonon damping rate ($\sim q^{z_b}$) at low-temperatures, with $z_b=3$ ($\Omega\sim q^3$) behavior arising far from the FFS fillings, and $z_b=5$ ($\Omega\sim q^5$) behavior occurring \emph{at} the FFS. Close to, but not precisely at the FFS fillings, there is an finite-temperature crossover (see Fig.~\ref{fig:schematic}) between $z_b=5$ dynamics at intermediate scales to $z_b=3$ at sufficiently low temperature scales where the phonons ``notice" the Fermi-surface curvature, with crossover scale:
\begin{align}
E_* \approx \frac{c_s^2 q_*^3}{\gamma}.
\end{align}

Furthermore, at sufficiently high energies or temperatures $E,T\gg E_\text{LD}$, the phonon-damping becomes unimportant, and the phonons become sharp (underdamped) quasi-particles, where 
\begin{align}
E_\text{LD}({z_b}) \approx
	\begin{cases}
		\sqrt{c_s \gamma}, & z_b=3 \\
		\(E_*^2 c_s^5 \gamma^5\)^{\frac{1}{12}}, & z_b=5
	\end{cases}
\end{align} 
For temperatures above $E_\text{LD}$, the phonons recover their ordinary $z_b=1$ ($\Omega\sim q$) dynamics.

The various dynamical scaling crossovers are depicted in Fig.~\ref{fig:schematic}, and summarized by:
\begin{align}
	z_b = 
	\begin{cases}
		1, & E>E_\text{LD} \\
		5, & E_*< E<E_\text{LD}\\ 
		3, & E<E_*,E_\text{LD}
	\end{cases}
\end{align}
For our numerical estimates of $E_\text{LD}$ below, we give results for the $z_b=3$ regime only, since the $z_b=5$ behavior arises only in a very narrow window around the FFS fillings where $m > \frac{1}{\Gamma} \sqrt{\frac{\pi \rho v_F c_s}{2\lambda}}$.

\begin{figure*}[t]
	\includegraphics[width=\textwidth]{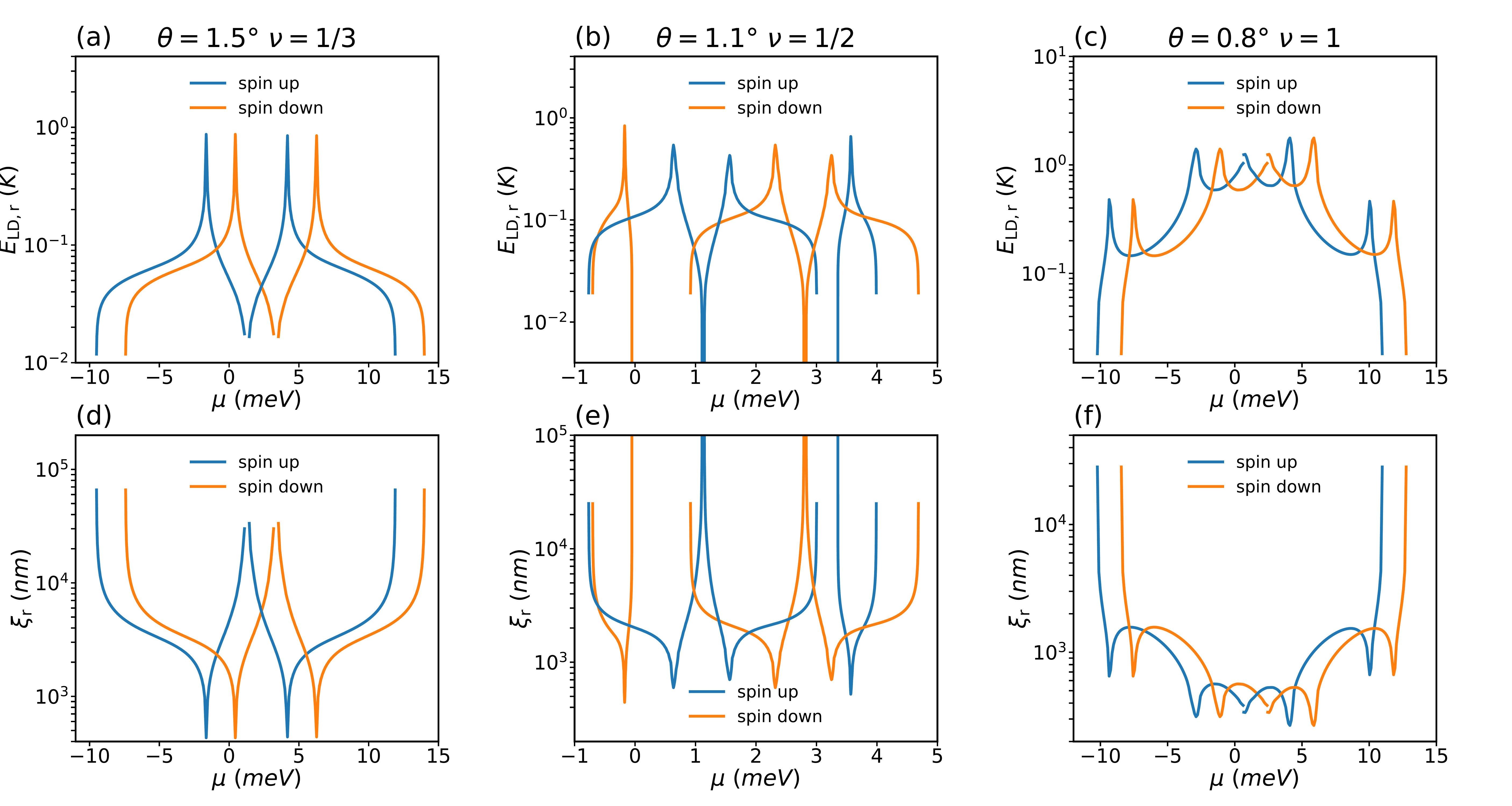}
	\caption{{\bf Landau damping energy and SAW attenuation length -- } (a), (b) Landau damping scale of relative-displacement phonon $E_{\rm LD,r}$ along path $K-\Gamma-K'$ for twist angle $\theta=1.5^\circ$, $\theta=1.1^\circ$ and flux filling $\nu=1/3$, $\nu=1/2$. (c) $E_{\rm LD,r}$ along  path $M-\Gamma-M'$ for twist angle $\theta=0.8^\circ$ and flux filling $\nu=1$. (d),(e) SAW attenuation length of relative-displacement phonon $\xi_{\rm r}$ along path $K-\Gamma-K'$, for twist angle $\theta=1.5^\circ$, $\theta=1.1^\circ$ and flux filling $\nu=1/3$, $\nu=1/2$ and typical driving frequency $\omega/2\pi=200{\rm MHz}\sim0.96\times10^{-2}{\rm K}$. (f) $\xi_{\rm r}$ along path $M-\Gamma-M'$, for twist angle $\theta=0.8^\circ$, filling $\nu=1$ and driving frequency $\omega/2\pi=200{\rm MHz}$}
	\label{fig:ELD} 
	\vspace{-0.2in}
\end{figure*}

The crossover scale, $E_\text{LD}$, will play an important role in our following discussion, and numerical results are shown in Fig.~\ref{fig:ELD} for representative parameters. 
$E_\text{LD}$ is generally deeply sub-Kelvin
except near FFS fillings where the vanishing Fermi-surface curvature results in divergent effective mass and Landau damping coefficient for select directions along the Fermi-surface. 
This divergence will be rounded in practice by disorder and higher-order terms in the electron dispersion, as shown in Fig.~\ref{fig:schematic}.
For $\theta=0.8^\circ$, $E_\text{LD}/k_B$ exceeds $1K$ over appreciable ranges of chemical potential ($\pm 0.3$meV) and density ($\pm 10^{-13}\text{cm}^{-2}$).

\subsection{Anomalous Surface Acoustic Wave Attenuation Signatures}
This Landau-damped form can be probed by surface acoustic wave (SAW) attenuation experiments.
SAWs can be injected and detected by piezoelectric contacts, which we model as semi-infinite in the $y$-direction, and separated by finite distance $L$ in the $x$ direction. We extract the SAW attenuation length, $\xi$, by computing the fixed-frequency phonon propagator between source to detector: $\int dy D(\omega,x=L,y)\sim e^{-L/\xi}$, and find (see Appendix \ref{app:saw}):
\begin{align}
	\xi = 
	\begin{cases} 
		2\omega/\gamma & \omega \gg E_\text{LD} \\
		\frac{2c_s}{E_\text{LD}}\(\frac{E_\text{LD}}{\omega}\)^{1/3} & \omega \ll E_\text{LD}
	\end{cases},
	\label{eq:xi}
\end{align}
with numerical results shown in Fig.~\ref{fig:ELD}. The FFS points near VH-fillings produce a singular suppression of SAW propagation, providing a characteristic fingerprint of the Landau damping in the doping-dependence of $\xi$.

\subsection{Transport properties}
Strong coupling to MH acoustic (MHA)-phonons is expected to result in unconventional non-Fermi liquid scaling for transport and thermodynamic properties with temperature and frequency~\cite{Lee1989,lee1992gauge,polchinski1994low,nayak1994non,Halperin1993,altshuler1994low}. A one-loop analysis predicts that, below $E_\text{LD}$, scattering by overdamped phonons with dynamical critical exponent $z_b$ results in (retarded) electron self-energy 
\begin{align}
\Sigma^R_f(\omega;z_b) = i \(E_{\rm NFL}(z_b)\)^{\frac{1}{z_b}} |\omega|^{\frac{(z_b-1)}{z_b}}
\end{align}
where
\begin{equation}
E_{\rm NFL}(z_b)= E_* \(\frac{q_*^2}{(z_b-1)\sin \(\pi/z_b\)m E_*}\)^{z_b}
\end{equation}
For asymptotically low energies ($E \ll E_{\rm NFL} \equiv {\displaystyle \text{min}_{z_b=3,5}} E_{\rm NFL}(z_b)$), the phonon-scattering self-energy dominates over the bare electron energy, resulting in a destruction of Fermi liquid quasi-particles. The electron dynamical critical exponent gets modified from $z_f=1$ to $z_f = \frac{z_b}{z_b-1}$. However, in TBG, this occurs only at inaccessibly low temperatures (in Appendix \ref{app:nflscales}, we estimate $E_{\rm NFL}/k_B\leq 10^{-4}\text{K}$ over the range of parameters explored in Fig.~\ref{fig:bands}). Therefore, in the experimentally accessible regime, the system will be in a ``precursor" NFL state, where electronic quasi-particles are ailing but not yet expired. 

Non-Fermi liquid behavior is often characterized through scaling of resistivity with temperature. The precursor NFL regime is divided into high-, intermediate-, and low-temperature regimes by  two important cross-over scales: i) the Bloch-Gruneisen (BG) temperature, $T_\text{BG}$ above which thermally activated phonons carry average momentum that is larger than $2k_F$ so that phonon-contributions to resistivity are dominated by large-momentum scattering by thermal excitations, and ii) the Landau-damping temperature $T_\text{LD}=E_\text{LD}/k_B$, below which the phonons become overdamped and non-Fermi liquid temperature dependence arises. Using the Born-approximation to compute the transport time, the asymptotic temperature dependence of the electron-phonon contributions to resistivity in each regime are summarized as follows:
\begin{align}
	\rho(T) \sim 
	\begin{cases}
		T; & T>T_\text{BG} \\
		T^2; & T_\text{BG}> T>T_\text{LD}\\ 
		T^{4/3}; & T<T_\text{LD}
	\end{cases}.
\end{align}
The phonon-mechanism for linear-$T$ resistivity in the high-temperature regime $(T>T_\text{BG})$ is a standard effect that does not rely on Moire-Hofstadter physics and can be observed in the absence of a $B$-field~\cite{hwang2008acoustic,Wu2019}.

By contrast, the phonon-induced $\rho(T)\sim T^2$ behavior in the intermediate regime arises from the singular coupling between electrons and MHA-phonons, and crucially relies on the magnetic field producing non-vanishing electron-phonon coupling vertex in the limit of $q\rightarrow 0$ (see Appendix~\ref{app:transport_rate} for detailed calculation and numerical results). This produces a strong enhancement of the electron-phonon-scattering contribution to resistivity compared to the conventional $T^4$ behavior~\cite{hwang2008acoustic} expected in the absence of a magnetic field. We remark that a different $T^2$-contribution to resistivity arises from electron-electron interactions in a Fermi-liquid, and a quantitative and perhaps model-dependent comparison will be required to distinguish the origin of this $\rho(T)\sim T^2$ scaling. For example electron-electron interaction contributions to resistivity may be neglected in very clean samples when Umklapp processes required to relax momentum are strongly suppressed by Fermi-surface geometry and momentum conservation requirements. Furthermore, as we elaborate in the next section, the enhanced electron-MHA-phonon scattering can have a an impact on the scaling of quantum oscillatory and cyclotron resonance phenomena that is qualitatively different than standard electron-electron or electron-phonon behavior.

Finally, in the low temperature regimes for $T<T_\text{LD}$, scattering from overdamped phonons results in departures from Fermi liquid behavior, whose properties can only be computed in artificial limits~\cite{Mross2010,Dalidovich2013}. For example, a potentially oversimplified scattering-rate calculation (neglecting hydrodynamic effects and umklapp scattering) predicts temperature-dependent resistivity $\rho(T)\sim T^{4/3}$ for $T<E_\text{LD}$~\cite{Lee1989,Ioffe1990}. We note that while flat-patches experience a larger scattering rate $\tau_\text{tr,FFS}^{-1}\sim T^{8/5}$, their contribution to transport is likely shorted-out by the coexisting curved patches that have lower $\sim T^{4/3}$ scattering. Other signatures include spatial decay of quasi-particle interference patterns, which expose the anomalous dimension of the electrons in the NFL state~\cite{Metlitski2010,Mross2010}.

\subsection{MHA Polarons}
Even for energy and temperature scales above $E_\text{LD}$, electrons and MHA-phonons are strongly coupled into unusual polaronic degrees of freedom, which we dub ``MHA-polarons". The dynamics of MHA polarons differs dramatically from their counter-parts in ordinary metals, since accelerating MHA polaron motion results in radiation of soft-collinear phonons in direct analogy to cyclotron radiation. In the intermediate-temperature regime, $T_\text{BG}>T>T_\text{LD}$, a tree-level scattering rate calculation using the undamped phonon propagator (see Appendix~\ref{app:transport_rate}) shows that the rate of phonon emission is logarithmically divergent in the infrared (though the energy emitted is finite), in analogy to soft collinear divergences in quantum electrodynamics. We analyze the MHA-polaron motion and phonon-radiation within a semiclassical framework (see Appendix~ref{app:polarons}), which amounts to an infinite-order resummation of soft collinear divergences in the eikonal approximation~\cite{laenen2009path}.

To physically probe the dynamics of the MHA-polaron, we consider detuning the magnetic field slightly away from commensurate filling $B = B_\nu +\Delta B$, with $e\Delta B\ll 1/|\vec{a}_{1,2}|^2$. Without phonons, electrons would exhibit cyclotron motion with effective cyclotron frequency, $\omega_{c,0} = \frac{e\Delta B}{m}$ set by the detuning field $\Delta B$ rather than the full field $B$ (as can be seen by semiclassical motion of a wave-packet made from Bloch states of the bands at commensurate filling $B_\nu$ moving in effective field, $\Delta B$). Oscillations with the reduced field $\Delta B$, rather than the full field $B$ could be measured by standard Shubnikov-de-Haas, cyclotron resonance, or current-focusing~\cite{Taychatanapat2013} techniques, but to our knowledge have not yet been explored. 

\subsubsection{Cyclotron motion}
Numerical computation reveals the electron-phonon coupling is proportional to the band-velocity, so that the phonon fields couple to the electron velocity, $\dot{\v{r}}$, where $\v{r}$ is the electron coordinate. This velocity coupling is directly analogous to the coupling between charged-particles and gauge fields. Incorporating this feature into a Schwinger-Keldysh path integral description of a single-electron (or hole) coupled to a zero-temperature bath of phonons, we obtain an effective action $\int_{\mathcal{C}} dt L[\v{r},\v{u}]$ on the closed-time Keldysh contour $\mathcal{C}$ with the Lagrangian:
\begin{align}
\mathcal{L}[\v{r},\v{u}] &= \mathcal{L}_e[\v{r}] + \mathcal{L}_{e-ph}[\v{r},\v{u}(\v{r})] + \mathcal{L}_{ph}[\v{u}] \nonumber\\
\mathcal{L}_e[\v{r}] &= \frac{1}{2} m \dot{\v{r}}^2 - e \dot{\v{r}} \cdot \v{A}(\v{r}), \nonumber \\
\mathcal{L}_{ph}[\v{u}] &= \frac{\rho}{2} \[\(\d_t \v{u}\)^2-c_s^2\(\nabla \v{u}\)^2\],
\nonumber\\
\mathcal{L}_{e-ph}[\v{r},\v{u}(\v{r})] &= \alpha \dot{\v{r}} \cdot \v{u}(\v{r}).
\end{align}
Here, the vector potential $\v{A}(\v{r})$ corresponds to the excess magnetic field ($\Delta B = \nabla \times \v{A}(\v{r})$), and does not include the constant field used to reach the MH-regime, and $\alpha \equiv m \frac{\d\Gamma(k,q=0)}{\d k}$.

We now integrate out the phonon field by approximating the change of the electron position to be slow compared to the wavelengths of the phonons being integrated out~\cite{fisher1986ground}, which leads to a quadratic, non-local in time effective action for the MHA polaron coordinate $\v{r}$:
\begin{align}
	S_\text{eff} =
	S_0- \frac{\alpha^2}{2\pi \rho c_s^2} \int_{-\infty}^{\infty} dt \int_{t}^{\infty} dt' \frac{\v{r}^{q}(t)\cdot \dot{\v r}^{cl}(t')} {\(t-t'\)^2}  
\end{align}
where $r^{cl/q}$ are the so-called ``classical" and ``quantum" combinations of $r(t)$ on the two-time Keldysh contour, and $S_0$ is the bare action for an electron in the absence of phonon degrees of freedom.
 Varying this effective action~\cite{kamenev2011field}, we find the following semiclassical equation of motion (EOM) for $\v{r}$~(see Appendix \ref{app:polaron}): 

\begin{align}
\ddot{\v{r}}(t) = \frac{\v{F}_\text{ext}}{m} -g^2\int_{-\infty}^t dt' \frac{\dot{\v{r}}(t')}{(t-t')^2}
\label{eq:eom}
\end{align}
where 
$g = \sqrt{\frac{m}{2\pi \rho c_s^2}}\frac{\d\Gamma(k,q=0)}{\d k} $ 
is a dimensionless measure of electron-phonon coupling strength, and $\v{F}_\text{ext}$ represents the external force due to the excess magnetic field~\footnote{At non-zero temperature an additional stochastic force would arise from thermal phonon fluctuations.}. The second term represents the non-Markovian effects due to the gapless phonons. 

The cyclotron motion of MHA-polarons differs markedly from that of bare electrons as we now show by solving the polaron equation of motion in Eq.~(\ref{eq:eom}) for damped oscillatory cyclotron motion $\v{r}(t) = r_0 e^{-\Gamma_c t} \( \cos (\omega_{c} t) \hat{x} + \sin (\omega_{c} t) \hat{y}\)$, in an excess field $\Delta B$. Defining an effective frequency dependent mass $m_p(\omega_{c}) \equiv eB/\omega_{c}$ we find that the MHA-polaron exhibits a scale-dependent logarithmically diverging effective mass: $m_p(\omega_{c}) = m\(1 + g^2 \log \frac{\Lambda}{\omega_{c}}\)$. The cyclotron frequency scales with an unconventional power of $\Delta B$:
\begin{align}
	\omega_{c} = \omega_{c,0}\( \frac{\omega_{c,0}}{\Lambda}\)^{g^2} \sim \( \Delta B \)^{1+g^2}
\end{align}
and a $\Delta B$ independent decay rate $\Gamma_c = g^2\Lambda$.
Here $\omega_{c,0} = \frac{e\Delta B}{m}$ is the bare (non-interacting) electron cyclotron frequency of the MH-bands. 
For TBG we find that the $g^2 \lesssim 10^{-2}$ (even close to the VH-fillings), so these effects may be challenging to observe. We note, however, that these predictions equally apply to a variety of analog non-Fermi liquid systems~\cite{Halperin1993,Oganesyan2001,Metlitski2010,Xu2010,Bahri2015} where the polaron coupling constant would not be suppressed by the electron-ion mass ratio and could potentially give appreciable modifications to semiclassical electron motion.

\subsubsection{Quantum oscillations}
In addition to energy dissipation by phonon radiation, MHA-phonons cause characteristic dephasing of phase-sensitive measurements such as  periodic-in-$1/\Delta B$ quantum oscillations in density of states and resistivity. As electrons experience MHA-phonon fields $\v{u}$ as an effective electromagnetic vector potential, zero-point fluctuations of MHA-phonons give rise to quantum-fluctuating geometric (Aharonov-Bohm-like) phase,  $e^{i\alpha \int dt ~\dot{\v{r}}\cdot \v{u}(\v{r}(t))}$. Quantum fluctuations in the phonon fields results in random geometric-phase accumulation that suppress the quantum oscillation amplitude.

We compute the quantum oscillatory contribution to density of states,$N(\e)$, at energy $\e$ via a semiclassical sum of the return amplitude for multiple classical cyclotron orbits of the MHA-polaron, averaged over fluctuating geometric phases due to MHA-phonons~\cite{mirlin1996quasiclassical}:
\begin{align}
N(\e) &\approx -\frac{1}{\pi}\int_0^t dt~\text{Im}~G^R(r=r';t)e^{-i\e t}
	\nonumber\\ &\approx
	\<\sum_n e^{-i \frac{2\pi \e}{\omega_c}n}e^{i\alpha \int dt ~\dot{\v{r}}\cdot \v{u}(\v{r}(t))}\>_u
\end{align}
where $\< \ldots \>_u$ indicates an average over quantum fluctuations of $\v{u}$. We restrict our attention to zero temperature here, as finite temperature effects with $T\ll \omega_c$ do not effect the scaling form that we identify. These fluctuations are approximately quenched on the time-scale of the cyclotron motion since the electrons are fast compared to the phonons, i.e. we may approximately replace $\int dt ~\dot{\v{r}}\cdot \v{u}(\v{r}(t))$ by $\Phi(\v{u})$, the flux of a static effective magnetic field $b = \nabla \times \v{u}$ through the cyclotron orbit. Within this approximation, applying the Poisson summation formula (see Appendix \ref{app:dos}) yields.
\begin{align}
N(\e) &= \sum_n e^{-i \frac{2\pi \e}{\omega_c}n}e^{-i\frac12\alpha^2 n^2\<\Phi\Phi\>}
	\nonumber\\&=
	\sum_n A_c \exp\[-\frac{1}{2\Gamma^2}\(\e-n\omega_{c,0}\)^2 \]
\end{align}
The resulting oscillator contribution to the density of states takes the form of a comb of Gaussian peaks centered at integer multiples of the cyclotron frequency $\omega_{c,0}$ with peak-amplitude $A_c\sim |\Delta B|$ and width $\Gamma \sim \(\Delta B\)^{2g^2}$. This contrasts the usual $A_c \sim \sqrt{\Delta B}$ scaling expected from impurity scattering, providing a means to distinguish these two mechanisms.

\section{Outlook} We have explored moir\'e super-lattice structures in high magnetic fields as potential platforms for simulating non-Fermi liquid (NFL) physics, due to unusually singular electron-phonon coupling. While the asymptotic low-temperature NFL fixed point remains at inaccessibly low temperature, intermediate scale non-Fermi liquid precursor behavior and unconventional polaron dynamics can be observed at by tuning carrier density near a van-Hove singularity. 
At or very close to the van Hove filling, the NFL physics will compete (or intertwine) with enhanced tendency to order with a nested Fermi surface. This competition appears to be controllable by twist-engineering~\cite{hsu2020topological,lin2019chiral,classen2019competing,chichinadze2019nematic,isobe2018unconventional,yuan2019magic}, and a more detailed study of the interplay between exotic interaction-driven orders, and non-Fermi liquid physics in these systems will be a compelling subject for future investigation.

\vspace{4pt}
\noindent{\it Acknowledgements -- } We thank Allan Macdonald, Elaine Li, Sid Parameswaran, Brad Ramshaw, Isabelle Phinney, Igor Blinov and Naichao Hu for insightful discussions. This work was supported by NSF DMR-1653007 (AP), and by the NSF through the Center for Dynamics and Control of Materials: an NSF MRSEC under Cooperative Agreement No. DMR-1720595 (AK). Part of this work was performed at the Aspen Center for Physics, which is supported by NSF grant PHY-1607611. 

\bibliography{MoireFractalNFL}

\appendix
\onecolumngrid

\section{Moire-Hofstadter bands\label{app:mhbands}}
In this section, we review the derivation of the magnetic Bloch bands for twisted bilayer graphene. Define the magnetic lattice translational operators:
\begin{equation}
\mathcal{T}_i=e^{i\v{K}\cdot\v{a}_i},
\end{equation}
where $i=1,2$, which perform translation by Moire lattice-vectors: $\v{a}_1=\frac{a}{\theta}\(-\frac{\sqrt{3}}{2},\frac{1}{2}\)$ and $\v{a}_2=\frac{a}{\theta}\(\frac{\sqrt{3}}{2},\frac{1}{2}\)$. Further, define guiding-center momenta:
\begin{equation}
K_x=\Pi_x+\frac{y}{l_B^2},\quad K_y=\Pi_y-\frac{x}{l_B^2},
\end{equation}
which satisfy $[K_x,K_y]=\frac{i}{l_B^2},\ [K_\alpha,\Pi_\beta]=0$, and $[\mathcal{T}_i,H]=0$. 

Unlike ordinary lattice-translations, magnetic translations do not generically commute except at special commensurate magnetic fields. Generically, 
\begin{equation}
\mathcal{T}_1\mathcal{T}_2=\mathcal{T}_2\mathcal{T}_1\exp\(i\frac{\sqrt{3}a^2}{2\theta^2l_B^2}\), 
\end{equation}
which vanishes if and only if:
\begin{equation}
\frac{\sqrt{3}a^2}{2\theta^2l_B^2}=2\pi\frac{p}{q}\quad{\rm or}\quad k_\theta^2l_B^2=\frac{4\pi }{3\sqrt{3}}\frac{q}{p},
\end{equation}
which gives the condition for commensuration of the Moire lattice and magnetic field.

At commensurate filling, $[\mathcal{T}_1,\mathcal{T}^q_2]=0$, so that we can construct the magnetic Bloch states $\vert\alpha\v{k}\rangle$ with band index $\alpha$, as the simultaneous eigenstates of $H$, $\mathcal{T}_1$, and $\mathcal{T}^q_2$:
\begin{align}
&H\vert\alpha\v{k}\rangle=E_{\alpha}(\v{k})\vert\alpha\v{k}\rangle, \\
&\mathcal{T}_1\vert\alpha\v{k}\rangle=e^{i\v{k}\cdot\v{a}_1}\vert\alpha\v{k}\rangle, \\
&\mathcal{T}^q_2\vert\alpha\v{k}\rangle=e^{iq\v{k}\cdot\v{a}_2}\vert\alpha\v{k}\rangle.
\end{align}
An extended magnetic Brillouin zone (MBZ) is spanned by magnetic-reciprocal lattice vectors: $\v{g}_1$ and $\v{g}_2/q$ (where $\v{g}_i$ are the original Moire reciprocal lattice vectors). As we now show, the commensurability integers $(p,q)$, determine the number of sub-bands, and size of the reduced MBZ. The commutation relations for $\mathcal{T}_1$ and $\mathcal{T}_2$ give:
\begin{equation}\label{Degeneracy}
\mathcal{T}_1\mathcal{T}^j_2\vert\alpha\v{k}\rangle=e^{i\(\v{k}+\v{g}_1jp/q\)\cdot\v{a}_1}\mathcal{T}^j_2\vert\alpha\v{k}\rangle
\end{equation}
implying a set of degenerate states  $\mathcal{T}^j_2\vert\alpha\v{k}\rangle\sim  \vert\alpha,\v{k}+\frac{jp}{q}\v{g}_1\rangle$ ($j=1,2,\dots,q-1$), which have identical energy since $[H,\mathcal{T}(\v{a}_2)]=0$. Thus there is a $q$-fold degeneracy along the $\v{g}_1$ direction. Equivalently, we can consider a reduced magnetic Brillouin zone spanned by $\v{g}_1/q$ and $\v{g}_2/q$ with $q$-fold degeneracy. In addition, Eq.(\ref{Degeneracy}) implies: $E_{\alpha}(\v{k})=E_{\alpha}\(\v{k}+\frac{p}{q}\v{g}_1\)$, that is, one period in a magnetic band extends over $p$ MBZ's in the direction of $\v{g}_1$. The reduction of energy bands to one MBZ will yield $p$ different subbands. For convenience, we extend $E_\alpha({\v{k}})$ to $E_\alpha(\v{k}+l\v{g}_1)\equiv E_{\alpha l}({\v{k}})$ where $l=0,1,\dots,p-1$ represents subbands and $\v{k}$ is restricted to the reduced MBZ.

Having identified the distinct magnetic sub-bands and reduced MBZ, we can label states of TBG by: $\vert \tau, \sigma, n,l,\v{k}\rangle$, where $\tau=1, 2$ represents layers, $\sigma=A, B$ represents sublattices, $n$ represents LL index, and $l$ represents subbands. For completeness we write the wavefunction of the LL states in the Landau gauge: $\< \v{r}| n, \v{k}\> \sim e^{i k_yy} H_n(x + k_y\l_B^2) e^{-{\(x + k_y\l_B^2\)}/2 l_B^2}$. However, the expressions obtained below are general and do not rely on fixing a particular gauge.
Under this basis, the single-layer Hamiltonian reads
\begin{equation}\label{singlelayerHamiltonian}
h(\theta/2)=\frac{\sqrt{2}v}{l_B}\(e^{i\theta/2}\sqrt{n+1}\vert 2, B, n+1, l,\v{k}\rangle\langle 2, A, n, l,\v{k}\vert+ h.c.\)
\end{equation}
To calculate matrix elements of the interlayer hopping term under this basis, we first split $e^{-i\v{g}_j\cdot\v{r}}$ as
\begin{equation}
e^{-i\v{g}_j\cdot\v{r}}=e^{-i\v{g}_j\cdot\v{\eta}}e^{-i\v{g}_j\cdot\v{R}}.
\end{equation}
where 
\begin{equation}
\v{R}=\begin{pmatrix}
R_x\\
R_y
\end{pmatrix}=\begin{pmatrix}
x-\Pi_yl_B^2\\
y+\Pi_xl_B^2
\end{pmatrix}\equiv\begin{pmatrix}
x-\eta_x\\
y-\eta_y
\end{pmatrix}
=\v{r}-\v{\eta}
\end{equation}
Notice that $\v{R}=l_B^2\(\hat{\v{z}}\times\v{K}\)$ only acts on $\vert l,\v{k}\rangle$ and $\v{\eta}=-l_B^2\(\hat{\v{z}}\times\v{\Pi}\)$ only acts on $\vert n \rangle$, so we have
\begin{equation}\label{tunnerlingterm}
\langle n,l,\v{k}\vert e^{-i\v{g}_j\cdot\v{r}}\vert n', l', \v{k}\rangle =\langle n\vert e^{-i\v{g}_j\cdot\v{\eta}}\vert n'\rangle\langle l,\v{k}\vert e^{-i\v{g}_j\cdot\v{R}}\vert l', \v{k}\rangle.
\end{equation}
The first term on the right hand side of Eq.(\ref{tunnerlingterm}) is given by $\langle n\vert e^{-i\v{g}_j\cdot\v{\eta}}\vert n'\rangle=F_{nn'}\(\v{g}_jl_B/\sqrt{2}\)$, where
\begin{equation}
F_{nn'}(\v{z})=\left\{\begin{aligned}&\sqrt{\frac{n'!}{n!}}(-z_x+iz_y)^{n-n'}e^{-\frac{z^2}{2}}\mathcal{L}_{n'}^{n-n'}(z^2)\ n\geqslant n'\\&\sqrt{\frac{n!}{n'!}}(z_x+iz_y)^{n'-n}e^{-\frac{z^2}{2}}\mathcal{L}_n^{n'-n}(z^2)\ n< n'\end{aligned}\right.
\end{equation}
with $\mathcal{L}$ being the associated Laguerre polynomial.

The second term on the right hand side of Eq.(\ref{tunnerlingterm}) can be converted to matrix elements of magnetic translation operators by using $e^{-i\v{g}_j\cdot\v{R}}=e^{-il_B^2\v{K}\cdot(\v{g}_j\times\hat{\v{z}})}$. Specifically:
\begin{align}
e^{-i\v{g}_1\cdot\v{R}}=\mathcal{T}^{-\frac{q}{p}}_2\quad{\rm and}\quad	e^{-i\v{g}_2\cdot\v{R}}=\mathcal{T}^{\frac{q}{p}}_1.
\end{align}
Considering $\mathcal{T}^{-\frac{q}{p}}_2\vert\alpha\v{k}\rangle\sim  \vert\alpha,\v{k}-\v{g}_1\rangle$ and applying the commutation relation between $\mathcal{T}_1$ and $\mathcal{T}_2$, we can obtain
\begin{align}
\mathcal{T}^{-\frac{q}{p}}_2&\vert l+1,\v{k}\rangle=e^{-i\frac{q}{p}\v{k}\cdot\v{a}_2}\vert l,\v{k}\rangle.\\
\mathcal{T}^{\frac{q}{p}}_1&\vert l,\v{k}\rangle=e^{i\frac{q}{p}(\v{k}\cdot\v{a}_1+2\pi l)}\vert l,\v{k}\rangle
\end{align} 
Hence, the interlayer hopping term in the MBZ reads
\begin{align}
T&(\v{k})=wT_0\vert2,\sigma,n,l,\v{k}\rangle\langle1,\sigma',n',l,\v{k}\vert\nonumber\\&+wT_1F_{nn'}\(\v{g}_1l_B/\sqrt{2}\)e^{-i\frac{q}{p}\v{k}\cdot\v{a}_2}\vert2,\sigma,n,l+1,\v{k}\rangle\langle1,\sigma',n',l,\v{k}\vert\nonumber\\&+wT_2F_{nn'}\(\v{g}_2l_B/\sqrt{2}\)e^{i\frac{q}{p}(\v{k}\cdot\v{a}_1+2\pi l)}\vert2,\sigma,n,l,\v{k}\rangle\langle1,\sigma',n',l,\v{k}\vert.
\end{align}
The numerical plots shown in the main text were obtained by diagonalizing this inter-layer hopping matrix truncated to a sufficiently large range of Landau indices $n$ to achieve well-converged cutoff-independent results. Notice that the direct truncation in LL basis will introduce an artifical zero mode in Eq.(\ref{singlelayerHamiltonian}). To remove it, we add a penalty term $E_P\vert2,A,N_{\rm cutoff}\rangle\langle2,A,N_{\rm cutoff}\vert$ with $E_P\gg\frac{\sqrt{2N_{\rm cutoff}}v}{l_B}$ to Eq.(\ref{singlelayerHamiltonian}) in the numerical calculation.

\section{MHA electron-phonon vertex}
In this section, we derive the MHA electron-phonon coupling.
For the convenience, we can write Hamiltonian under the lattice distortion $H_{\v{u}}$ as $
H_{\v{u}}=H_0(\v{\Pi})+H_{\rm t}(\v{r}-\v{u}(\v{r}))$,
where
\begin{equation}
H_0(\v{\Pi})=\begin{pmatrix}
h(-\theta/2)&0\\
0& h(\theta/2)
\end{pmatrix}
\end{equation}
and
\begin{equation}
H_{\rm t}(\v{r}-\v{u}(\v{r}))=\begin{pmatrix}
0& T(\v{r}-\v{u}(\v{r}))\\
T^\dagger(\v{r}-\v{u}(\v{r}))&0
\end{pmatrix}.
\end{equation}
Then we implement the co-moving frame transformation with $W_{\v{u}}=e^{-i\v{u}(\v{R})\cdot\v{\Pi}}$ on $H_0(\v{\Pi})$ and $H_{\rm t}(\v{r}-\v{u}(\v{r}))$ respectively. Since $[\v{R},\v{\Pi}]=0$, for small $\v{u}$ we have
\begin{equation}
W^\dagger_{\v{u}}\Pi_i W_{\v{u}}=\Pi_i+i[\v{u}(\v{R})\cdot\v{\Pi},\Pi_i]=\Pi_i-eB\e_{ij}u_j(\v{R})+\mathcal{O}(u^2)
\end{equation}
and corresponding transformed $H_0(\v{\Pi})$:
\begin{align}
W^\dagger_{\v{u}}H_0(\v{\Pi})W_{\v{u}}&=H_0(\v{\Pi}-e\v{B}\times\v{u}(\v{R}))\\=H_0(\v{\Pi})+&ve\v{u}\cdot\begin{pmatrix}
\v{B}\times\v{\sigma}_{-\theta/2}& 0\\
0&\v{B}\times\v{\sigma}_{\theta/2}
\end{pmatrix}+\mathcal{O}(u^2)
\end{align}
where $\v{\sigma}_{\theta/2}\equiv\v{R}_{\theta/2}\v{\sigma}$ are rotated Pauli matrices.

Similarly, for the tunneling term, we have
\begin{equation}
W^\dagger_{\v{u}}T(\v{r}-\v{u}(\v{r}))W_{\v{u}}=T(\v{r})+\mathcal{O}(\nabla\v{u},u^2)
\end{equation}
and therefore the transformed total Hamiltonian is
\begin{equation}
W^\dagger_{\v{u}}H_{\v{u}}W_{\v{u}}=H_{\v{u}=0}+ve\v{u}\cdot\begin{pmatrix}
\v{B}\times\v{\sigma}_{-\theta/2}& 0\\
0&\v{B}\times\v{\sigma}_{\theta/2}
\end{pmatrix}+\mathcal{O}(\nabla\v{u},u^2).
\end{equation}
which gives electron-phonon coupling for both the mean displacement and the relative displacement in the transformed coordinate:
\begin{equation}
{H}_{\rm e-ph}=\bar{\v{u}}\cdot\v{\Gamma}_{\rm m}+\v{d}\cdot\v{\Gamma}_{\rm r}
\end{equation}
with
\begin{equation}
\v{\Gamma}_{\rm m}=veB\begin{pmatrix}
\hat{\v{z}}\times\v{\sigma}_{-\theta/2}& 0\\
0&\hat{\v{z}}\times\v{\sigma}_{\theta/2}
\end{pmatrix}
\end{equation}
and
\begin{equation}
\v{\Gamma}_{\rm r}=-\frac{veB}{2\tan\frac{\theta}{2}}\begin{pmatrix}
\v{\sigma}_{-\theta/2}& 0\\
0&\v{\sigma}_{\theta/2}
\end{pmatrix}
\end{equation}
\section{Computation of non-Fermi liquid scales \label{app:nflscales}}
In this section, we estimate the relevant energy scales for the onset of Landau damping ($z_b=3$) for the phonons, and non-Fermi liquid behavior of electrons, using self-consistent one-loop propagators (random phase approximation).  While uncontrolled at asympotitically low-temperatures in the non-Fermi liquid regime, this approximation produces a reliable estimate of the onset for non-Fermi liquid behavior approaching from the high-temperature perturbative regime. Unless otherwise specified, all propagators are specified in imaginary time (Matsubara frequency).

\subsection{One-loop self-energies and propagators}
The (imaginary time/Matsubara) bare phonon propagator and bare electron propagator are given respectively by
\begin{equation}
D_0^{-1}(\v{q},\Omega)=-\rho_{\lambda} \(\Omega^2+\omega^2_{\v{q}}\)
\end{equation}
\begin{equation}
G_0^{-1}(\v{k},\omega)=i\omega-\e_{\v{k}}
\end{equation}
where for the long wave limit the phonon dispersion is $\omega_{\v{q}}\approx c_s\vert q\vert$ and $\e_{\v{k}}$ is the electron dispersion and $\lambda={\rm m}/{\rm r}$ is the label for mean and relative displacement phonon modes respectively.

With electron-phonon interaction, the full phonon propagator depends on the phonon self-energy as $D^{-1}=D_0^{-1}-\Pi$. At the lowest order of electron-phonon interaction, the one-loop phonon self-energy can be calculated by
\begin{equation}
\Pi(\v{q},\Omega)=\int\frac{d^2kd\omega}{(2\pi)^3}G_0(\v{k},\omega)G_0(\v{k}+\v{q},\omega+\Omega)\Gamma^2_{\alpha,\lambda}(\v{k},\v{q})
\end{equation}
where $\alpha={\rm L}/{\rm T}$ is the label for longitudinal and transverse mode, $\rho_{\rm m}=2\rho$, $\rho_{\rm r}=\frac{\rho}{2}$ ($\rho$ is the mass density of single layer graphene).
At the limit ${q}\rightarrow0$, only the nonvanished coupling  $\Gamma^2_{\alpha,\lambda}(\v{k},\hat{{q}})$ is left. After performing the integrals for the patch dispersion $\e_{\v{k}}=v_Fk_{\perp}+\frac{k_\parallel^2}{2m}$, we can obtain
\begin{equation}
\Pi(\v{q},\Omega)=\gamma_{\alpha,\lambda}\frac{\vert\Omega\vert}{\vert q_{\parallel}\vert}
\end{equation}
where $q_{\parallel}$ is the component parallel to the fermi surface and the Landau damping coefficient $\gamma_{\alpha,\lambda}$ is given by
\begin{equation}
\gamma_{\alpha,\lambda}(\hat{\v{q}})=\frac{\vert m\vert}{2\pi \rho_\lambda v_F}\left.\Gamma^2_{\alpha,\lambda}(\v{k},\hat{{q}})\right\vert_{\hat{\v{q}}\parallel {\rm FS}}
\end{equation}
where $m$ is the effective mass of electron. Then, the full phonon propagator reads
\begin{equation}
D(\v{q},\Omega)=-\frac{1}{\rho_{\lambda} \(\Omega^2+\omega^2_{\v{q}}+\gamma_{\alpha,\lambda}\frac{\vert\Omega\vert}{\vert q_{\parallel}\vert}\)}.
\end{equation}
Using $\Omega\sim c_s q$, the energy scale below which Landau damped modes dominate is
\begin{equation}
E_{\rm LD,\alpha,\lambda}\sim\sqrt{\gamma_{\alpha,\lambda}c_s}=\sqrt{\frac{c_s\vert m\vert}{2\pi \rho_\lambda v_F}\left.\Gamma^2_{\alpha,\lambda}(\v{k},\hat{{q}})\right\vert_{\hat{\v{q}}\parallel {\rm FS}}}
\end{equation}

On the other hand, using the Landau damped phonon propagator
\begin{equation}
D(\v{q},\Omega)\approx-\frac{1}{\rho_{\lambda}\(\omega^2_{\v{q}}+\gamma_{\alpha,\lambda}\frac{\vert\Omega\vert}{\vert q_{\parallel}\vert}\)},
\end{equation}
we can obtain the one-loop electron
self-energy at the Fermi level defined by $G^{-1}=G_0^{-1}-\Sigma$
\begin{align}
\Sigma(\omega)&=\int\frac{d^2qd\Omega}{(2\pi)^3}D(\v{q},\Omega)G_0(\v{q}+\v{k_{\rm F}},\Omega+\omega)\Gamma^2_{\alpha,\lambda}(\v{k},\hat{{q}})\nonumber\\&=i\(\frac{\Gamma^4_{\alpha,\lambda}}{12\sqrt{3}\pi^2\rho^2_{\lambda}v^2_{\rm F}\vert m\vert c_s^{4}}\)^{\frac{1}{3}}\vert\omega\vert^{\frac{2}{3}}{\rm sgn}(\omega)
\end{align}
which gives the Non-Fermi liquid energy scale
\begin{equation}
E_{\rm NFL}\sim \frac{\Gamma^4_{\alpha,\lambda}}{12\sqrt{3}\pi^2\rho^2_{\lambda}v^2_{\rm F}\vert m\vert c_s^{4}}
\end{equation}
below which the quasiparticle lifetime goes as $\tau^{-1}\propto\omega^{\frac{2}{3}}$.

\subsection{SAW attenuation length\label{app:saw}}
In this section, we give
the propagation intensity of a surface acoustic wave on twisted bilayer graphene, which is attenuated by the electron-phonon scattering and decaying as $e^{-r/\xi}$. Here we consider the surface acoustic wave generated by a line source and the phonon propagator in the direction perpendicular to the source is given by, the retarded phonon propagator:
\begin{equation}
D^R(r,\hat{\v{q}},\omega)=\frac{1}{\rho_{\lambda}}\int dqe^{iqr}\frac{1}{\omega^2-c_S^2q^2-i\gamma(\hat{\v{q}})\frac{\omega}{\vert q\vert}}
\end{equation}
(where we have analytically continued from Matsubara to retarded frequency: $i\Omega\rightarrow \omega+i0^+$).

For further convenience, we first define dimensionless quantities: $x\equiv qc_S/\omega$, $y\equiv r\omega/c_S$, and $\lambda\equiv E_{\rm LD}/\omega$ and then the integral becomes
\begin{align}\label{phononpropagator}
D^R&=\frac{2i}{\rho_{\lambda} c_s \omega}\int_0^\infty dx\frac{x\sin(xy)}{x-x^3-i\lambda^2}\nonumber\\&=\frac{2i}{\rho_{\lambda} c_s \omega}\int_0^\infty dx\frac{(x^2-x^4)\sin(xy)}{(x-x^3)^2+\lambda^4}-\frac{2}{\rho_{\lambda} c_s \omega}\int_0^\infty dx\frac{\lambda^2x\sin(xy)}{(x-x^3)^2+\lambda^4}
\end{align}

We focus on  the first term $I_1$ which is the imaginary part of $D$ and dictates the attenuation of SAWs
\begin{equation}
I_1=\frac{i}{\rho_{\lambda} c_s \omega}{\rm Im}\int^\infty_{-\infty}dx\frac{(x^2-x^4)e^{ixy}}{(x-x^3)^2+\lambda^4}
\end{equation} 
Note that for larger imaginary part of poles, the decaying rate of the integral is larger, thus we want to find the pole with the smallest positive imaginary part which dominates the attenuation. Solving $(x-x^3)^2+\lambda^4=0$ gives $6$ roots: $\beta\pm i\alpha$, $-\beta\pm i\alpha$ and $\pm i2\alpha$, where
\begin{equation}
\alpha=\frac{2^{\frac{1}{3}}\eta^{\frac{2}{3}}-2\cdot3^{\frac{1}{3}}}{2\cdot6^{\frac{2}{3}}\eta^{\frac{1}{3}}},\quad
\beta=\frac{2^{\frac{1}{3}}3^{\frac{1}{2}}\eta^{\frac{2}{3}}+2\cdot3^{\frac{5}{6}}}{2\cdot6^{\frac{2}{3}}\eta^{\frac{1}{3}}}
\end{equation}
with $\eta=\sqrt{12+81\lambda^4}+9\lambda^2$. Hence
the smallest positive imaginary part of poles is $\alpha$ and correspondingly the typical decaying length $\xi$ is given by
\begin{equation}
\xi=\frac{r}{y\alpha}=\frac{c_S\lambda}{E_{\rm LD}\alpha}
\end{equation}
When energy scale of acoustic wave is much larger than Landau damping scale, i.e., $\lambda\ll1$, $\alpha=\frac{\lambda^2}{2}+\mathcal{O}(\lambda^6)$, thus
\begin{equation}
\xi\approx\frac{2c_S}{E_{\rm LD}\lambda}=\frac{2\omega}{\gamma}
\end{equation}
On the other hand, for energy scale much smaller than Landau damping scale, i.e., $\lambda\gg1$, $\alpha=\frac{1}{2}\lambda^{\frac{2}{3}}+\mathcal{O}\(\lambda^{-\frac{2}{3}}\)$, thus
\begin{equation}
\xi\approx\frac{2c_S}{E_{\rm LD}}\(\frac{E_{\rm LD}}{\omega}\)^{\frac{1}{3}}
\end{equation}

\subsection{Perturbative electron lifetime in intermediate-temperature regime\label{app:transport_rate}}
To compute the electronic life-time for temperatures or frequencies above the Landau-damping scale $E_\text{LD}$, we compute the tree-level electron-phonon quantum and transport scattering rates using the bare (undamped) phonon propagator, and focus on the relative (interlayer-sliding) phonons which have two orders of magnitude stronger interactions with electrons. 

We work in a patch description of the Fermi-surface, which captures the universal aspects of the leading-order singularities in the electron-phonon scattering rate, while neglecting smooth, non-singular ``background" scattering processes. We consider an electron at initial momentum $\v{k}=k\hat{x}$ above the Fermi surface in a particular ($x$) direction (i.e. $\v{k}$ is measured relative to the Fermi-momentum $k_F\hat{x}$). We linearize the electron dispersion near the Fermi energy, and neglect the curvature of the Fermi-surface, $\e_{\v{p}} \approx v_F p_x$. In this computation, the electron dispersions dominates the energetics in the $x$ direction parallel to the Fermi-velocity, and the phonon dispersion controls the energetics in the $y$ (perpendicular) direction, allowing us to neglect the dispersion of the phonons along $x$: $\omega_{\v{q}}\approx c_s|q_y|$ (formally, this is justified for $c_s\ll v_F$). For tree-level scattering rates, we may consider the Fermi-sea as Pauli-blocked, since at this order in perturbation theory, additional virtual electron-hole pairs cannot be excited. These approximations capture the leading singular behavior of scattering rates in the perturbative intermediate-temperature scale regime.

Writing the relative layer displacement field $\v{d}$ in terms of annihilation operators $a_{\lambda,\v{q}}$ that destroy a phonon with polarization $\lambda$ and wave-vector $\v{q}$:
\begin{align}
\v{d} = \sum_{\lambda,\v{q}} \frac{1}{\sqrt{2\rho_\lambda c_s|\v{q}|}} \(\v{\eps}_{\lambda,\v{q}}a_{\lambda,\v{q}}e^{i\v{q}\cdot\v{r}}+h.c.\),
\end{align}
In the above-described setup, the electron couples only to a single polarization of the phonons, so we may drop the polarization indices, and write the corresponding electron-phonon coupling interaction coefficient as $\Gamma_{r,\lambda,\v{k}}:=\Gamma$

We first compute the total electron scattering-rate, which enters the so-called ``quantum" scattering rate that effects quantum-coherent processes such as quantum-oscillations:
\begin{align}
\tau_Q^{-1} = 2\pi \int_0^k \frac{dq_x}{2\pi} \int \frac{dq_y}{2\pi} \frac{\Gamma^2}{2\rho_r c_s|q_y|}\delta\(v_Fq_x-c_s|q_y|\) = \frac{\Gamma^2}{4\pi \rho_rc_sv_F}\int_0^k \frac{dq_x}{q_x}
\end{align}
where the $\delta$-function enforces energy conservation, and the integration limits $[0,k]$ on $q_x$ reflect momentum conservation and Pauli exclusion. The integral is logarithmically divergent in the infrared (IR, small $q$). This effect is directly analogous to similar infrared singularities arising in soft-photon emission in quantum electrodynamics (QED), where the logarithmic divergence is physically cutoff by an external IR scale such as the energy resolution of the detector. In this regime we see that the quantum lifetime is dominated by very small angle, low-energy scattering processes.  To properly account for these infra-red divergence, in the following section, we implement a semiclassical treatment which is equivalent to resumming the divergent soft-collinear radiation processes to infinite order~\cite{laenen2009path}.

\begin{figure*}[t]
	\includegraphics[width=\textwidth]{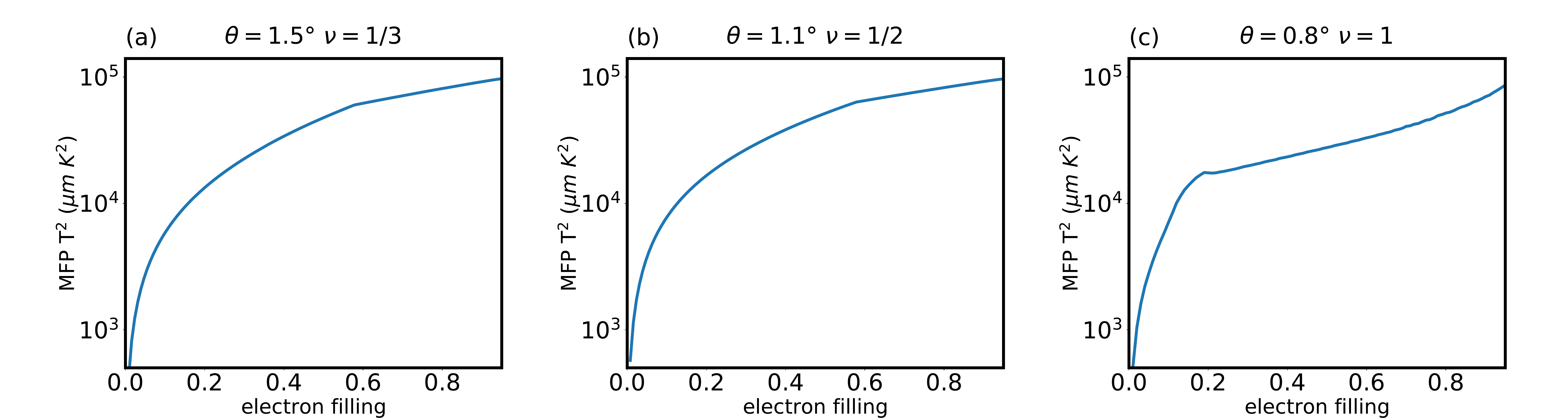}
	\caption{{\bf Prefactor of mean free path -- } $l\times T^2$ versus electron filling for (a) $\theta=1.5^\circ$, $\nu=1/3$ and (b) $\theta=1.1^\circ$, $\nu=1/2$ and (c) $\theta=0.8^\circ$, $\nu=1$.}
	\label{fig:mfp} 
	\vspace{-0.2in}
\end{figure*}

For transport, small angle scatterings are ineffective at relaxing momentum. A standard cheap way of accounting for this is to weight the scattering processes by the change in angle, which for small angle scattering can be approximated by inserting a factor of $\(\frac{q_y}{k_F}\)^2$ into the above integrals. This angle-weighting results in a non-diverging transport scattering rate:
\begin{align}
\tau_\text{tr}^{-1} = \frac{\Gamma^2v_F}{8\pi \rho_rc_s^3}\(\frac{k}{k_F}\)^2 \sim T^2
\end{align}
At temperature $T$, electrons typically have momentum $k\sim T/v_F$ relative to the Fermi surface, and we see that the above expression gives a $\rho(T)\sim T^2$ contribution to resistivity. Fig. \ref{fig:mfp} shows the prefactor of the mean free path $l\sim T^{-2}$ versus electron filling. We emphasize that, despite the Fermi-liquid like scaling of resistivity, the logarithmically divergent quantum-lifetime signals a breakdown of Fermi-liquid quasi-particles.

\section{MHA Polaron dynamics~\label{app:polarons}}
In this section, we study the problem of an electron directly coupled to MHA-phonons for intermediate to high energy regime where the phonons are not yet overdamped (the analogous problem in the asymptotic low-temperature non-Fermi liquid regime was previously studied using a quantum Boltzmann equation approach~\cite{kim1995quantum}). This situation is relevant not only for MH-polaron problems involving a single electron, but also allows a non-perturbative effective resummation of divergent radiative corrections due to emission of soft collinear phonons at energy scales exceeding $E_\text{LD}$. By numerical computation, we find that $\<n, \v{k}| \hat{\sigma}_i|n, \v{k} \> \approx \beta \frac{k_i}{k_{\theta}}$ for small $k_i/k_{\theta}$ with $\beta$ being a constant. The electron-phonon vertex is then, 
\begin{align}
\Gamma_{n,m,\lambda}(\v{k},\v{q}=0) = \frac{\beta ev_F B}{k_{\theta}} \hat{z} \times \v{k}, \quad
\Gamma_{n,r,\lambda}(\v{k},\v{q}=0) =  \frac{\beta ev_F B}{\theta k_{\theta}} \v{k}
\end{align}
Because of its enhanced coupling by a factor of $1/\theta$, we focus on the layer-antisymmetric phonon mode, in the following. The electron-phonon coupling, using first-quantized notation for the electron coordinate $\v{r}$ and momentum $\v{k}$, and second quantized for the phonon field $\v{u}$, and keeping only the leading term at small phonon momentum $q$ (dropping gradient coupling terms) is,
\begin{align}
{H}_\text{e-ph}(\v{r},\v{k}) &= \frac{\alpha}{m} \sum_{\v{q}} e^{i \v{q} \cdot \v{r}} \v{u}(\v{q}) \cdot \v{k} = \frac{\alpha}{m} \v{u}(\v{r}) \cdot \v{k}
\end{align}
with $\alpha = \frac{\beta e v_F B m}{\theta k_{\theta}}$. The coupling is analogous to the paramagnetic coupling of the particle to a $U(1)$ gauge field. Our problem is therefore, equivalent to that of a charged particle coupled minimally to a fluctuating gauge field. 

For simplicity, we consider the phonon spectrum to be isotropic with a polarization independent velocity, as it will not effect the universal physics. The effective Feynman Lagrangian for the particle in the presence of a uniform excess magnetic field discussed in the main text $\Delta \v{B} = \nabla \times \v{A}(\v{r})$ is then:
\begin{align}
L[\v{r},\v{u}] &= L_e[\v{r}] + L_{e-ph}[\v{r},\v{u}(\v{r})] + \int d^2r \mathcal{L}_{ph}[\v{u}] \nonumber\\
L_e[\v{r}] &= \frac{1}{2} m \dot{\v{r}}^2 - e \dot{\v{r}} \cdot \v{A}(\v{r}) \nonumber \\
\mathcal{L}_{ph}[\v{u}] &= \frac{\rho}{2} \[\(\d_t \v{u}\)^2-c_s^2\(\nabla \v{u}\)^2\]
\nonumber\\
L_{e-ph}[\v{r},\v{u}(\v{r})] &= \alpha \dot{\v{r}} \cdot \v{u}(\v{r})
\end{align}

The dynamics of the electron is naturally described in the Schwinger-Keldysh path integral formulation~\cite{kamenev2011field}. The corresponding action on the closed-time Keldysh contour $\mathcal{C}$ is:
\begin{align}
S[\v{r},\v{u}] = \int_{\mathcal{C}} dt L[\v{r},\v{u}]
\end{align}
Decomposing the fields $\v{u}(\v{r},t)$ and $\v{r}(t)$ in terms of fields residing on the forward time contour -- $\v{u^+}(\v{r},t)$ and $\v{r^+}(t)$, and the backward time contour -- $\v{u^-}(\v{r},t)$ and $\v{r^-}(t)$, we rewrite the action: 
\begin{align}
S[\v{r},\v{u}] = \int_{-\infty}^{\infty} dt \( L[\v{r^+},\v{u^+}] - L[\v{r^-},\v{u^-}] \)
\end{align}
We now perform standard rotation for the fields:
\begin{align}
\v{r}^{cl}(t) &= \frac{1}{2} \[\v{r}^+(t) + \v{r}^-(t) \], \quad \v{u}^{cl}(\v{r},t) = \frac{1}{2} \[\v{u}^{+}(\v{r},t) + \v{u}^{-}(\v{r},t) \] \nonumber \\
\v{r}^{q}(t) &= \frac{1}{2} \[\v{r}^+(t) - \v{r}^-(t) \], \quad \v{u}^{q}(\v{r},t) = \frac{1}{2} \[\v{u}^{+}(\v{r},t) - \v{u}^{-}(\v{r},t) \]
\end{align}
and obtain the following action: 
\begin{align}
S &= S_e[\v{r}] + S_{ph}[\v{u}] + S_{e-ph}[\v{r},\v{u}] \nonumber \\
S_e[\v{r}] &= \int_{-\infty}^{\infty} dt \[ -2m \v{r}^q \cdot \ddot{\v{r}}^{cl} - \( \dot{\v{r}}^{cl} + \dot{\v{r}}^{cl} \) \cdot \v{A}\( \v{r}^{cl} + \v{r}^{q} \) + \( \dot{\v{r}}^{cl} - \dot{\v{r}}^{cl} \) \cdot \v{A}\( \v{r}^{cl} - \v{r}^{q} \) \] \nonumber \\
S_{ph}[\v{u}] &= \frac{1}{2} \int_{-\infty}^{\infty} dt d^2r \vec{\v{u}}^T \hat{D}^{-1} \vec{\v{u}} \nonumber \\
S_{e-ph}[\v{r},\v{u}] &= \alpha \int_{-\infty}^{\infty} dt \begin{pmatrix} \v{u}^{cl}(\v{r}^{+}) + \v{u}^{q}(\v{r}^{+}) & \v{u}^{cl}(\v{r}^{-}) - \v{u}^{q}(\v{r}^{-})  \end{pmatrix}
\begin{pmatrix} 1 &1 \\ -1 &1 \end{pmatrix}
\begin{pmatrix} \dot{\v{r}}^{cl} \\ \dot{\v{r}}^{q} \end{pmatrix}
\end{align}
where we express the field $\v{u}$ by a vector in the Keldysh cl-q space:
\begin{align}
\vec{\v{u}}(t) = \begin{pmatrix} \v{u}^{cl}(t) \\ \v{u}^q(t) \end{pmatrix}, \quad
\hat{D}^{-1} = \begin{pmatrix} 0 & \[D^{-1}\]^A \\ \[D^{-1}\]^R & \[D^{-1}\]^K \end{pmatrix}
\end{align}
Here, $\frac{1}{2} \[ D^{-1} \]^{R(A)} = \rho \( \(\d_t \pm i0^+\)^2 - c_s^2 \nabla^2\)$. To study the motion of the electron coupled to the phonon bath, we first integrate out the phonons, and further obtain the classical equation of motion for the electron. Integrating out $\v{u}$, the effective action for the electron is,
\begin{align}
S_\text{eff}[\v{r}] &= S_e[\v{r}] + \frac{\alpha^2}{2} \int_{-\infty}^{\infty} dt dt'
\d_t\vec{\tilde{\v{r}}}(t)\nonumber\\
&\begin{pmatrix} (D^K + D^R + D^A)\(\v{r}^+(t)-\v{r}^+(t'),t-t'\) &(D^K - D^R + D^A)\(\v{r}^+(t)-\v{r}^-(t'),t-t'\) \\ (D^K + D^R - D^A)\(\v{r}^-(t)-\v{r}^+(t'),t-t'\) &(D^K - D^R - D^A)\(\v{r}^-(t)-\v{r}^-(t'),t-t'\) \end{pmatrix}
\d_{t'} \vec{\tilde{\v{r}}}(t')
\label{eq:action_ex}
\end{align}
where 
\begin{align}
\vec{\tilde{\v{r}}}(t) \equiv \begin{pmatrix} 1 &1 \\ -1 &1 \end{pmatrix}
\begin{pmatrix} \v{r}^{cl}(t) \\ \v{r}^q(t) \end{pmatrix}, \quad
\end{align}
We now make two simplifying approximations on the effective action. Following Ref.~\cite{fisher1986ground}, we are interested in electron motion where its position is changing slowly compared to the relevant wavelengths of the phonons being integrated out (Lamb-Dicke approximation). This corresponds to replacing $e^{-i\v{q} \cdot \Delta \v{r}} \approx 1$ in the following expression for the advanced and the retarded Green's function components:
\begin{align}
D^{R/A}(\Delta \v{r}, \Delta t) &= \frac{1}{\rho} \int \frac{d\Omega d^2q}{(2\pi)^3}\frac{e^{-i\v{q} \cdot \Delta \v{r}}e^{-i\Omega\Delta t}}{\(\Omega \pm i0^+\)^2-c_s^2q^2} \nonumber \\
&\approx \mp \frac{1}{2\pi \rho c_s^2} \frac{\theta(\pm \Delta t)}{\Delta t}
\end{align}
with $\Delta\vec{r}=\vec{r}'-\vec{r}$ and $\Delta t = t'-t$. The effective action [\ref{eq:action_ex}] is then quadratic in $\v{r}$:
\begin{align}
S_\text{eff} &= S_e[\v{r}] + \frac{\alpha^2}{2} \int_{-\infty}^{\infty} dt dt' \dot{\v{r}}(t)
\begin{pmatrix} 0 &D^A \\ D^R &D^K \end{pmatrix} \dot{\v{r}}(t')
\end{align}
Furthermore, we restrict ourselves to semiclassical dynamics of the electron at zero temperature; thus ignoring the role of thermally excited phonons. In this limit, we can ignore the Keldysh component of the Green's function $D^K$, in which case, the effective action takes the form:
\begin{align}
S_\text{eff} = S_e[\v{r}] + \frac{\alpha^2}{2\pi \rho c_s^2} \int_{-\infty}^{\infty} dt \int_{t}^{\infty} dt' \frac{\dot{\v{r}}^{cl}(t) \cdot \dot{\v{r}}^{q}(t')}{t-t'} 
\end{align}
Expanding the vector potential $\v{A}$ to first order in $\v{r}^q$, and integrating the non-local in time term by parts we get,
\begin{align}
S_\text{eff} = \int_{-\infty}^{\infty} dt \[ -2m \v{r}^q \cdot \ddot{\v{r}}^{cl} - 2 \v{r}^{q} \cdot \(\dot{r}_i^{cl} \nabla A_i(\v{r}^{cl}) -  \frac{d\v{A}(\v{r}^{cl})}{dt} \) - \frac{\alpha^2}{2\pi \rho c_s^2} \v{r}^{q}(t) \cdot \int_{t}^{\infty} dt' \frac{\dot{\v{r}}^{cl}(t')}{\(t-t'\)^2}  \]
\end{align}
\subsection{Polaron cyclotron motion\label{app:polaron}}
We obtain the classical equation of motion of the polaron by integrating over $\v{r}^q$, and expanding $\frac{d\v{A}(\v{r}^{cl})}{dt} = \dot{r}_i^{cl} \d_i \v{A}(\v{r}^{cl})$:
\begin{align}
\d^2_t \v{r}^{cl} = -\frac{e \v{\dot{r}}^{cl} \times \Delta \v{B}}{m} - g^2\int_{-\infty}^{t} dt' \frac{\dot{\v{r}}^{cl}(t')}{\(t-t'\)^2}
\label{eq:pol_eom}
\end{align}
where the dimensionless coupling constant
\begin{align}
g^2 \equiv \frac{\alpha^2}{2\pi \rho m c_s^2} = \frac{\beta^2}{8\sqrt{3}\pi} \frac{m}{M} \(\frac{v_F \nu}{c_s}\)^2
\end{align}
We now solve for the cyclotron motion of the polaron: $\v{r}(t) = r_0 \( \cos (\omega t) \hat{x} + \sin (\omega t) \hat{y}\)$ where we have suppressed the cl label, for convenience. We take into account damping effects by allowing $\omega$ to be complex. To solve for $\omega$, we first express $\v{r}(t)$ as a complex variable with the $x$ and $y$ components denoting the real and imaginary parts respectively. Now, using the polaron equation of motion (Eq.~\ref{eq:pol_eom}), we get:
\begin{align}
\omega = \omega_{c,0} - i g^2 \int_{t}^{\infty} dt' \frac{i e^{i\omega(t-t')}}{(t-t')^2}
\end{align}
where $\omega_{c,0} \equiv \frac{eB}{m}$. Defining an effective frequency-dependent mass for the polaron: $m_p(\omega) \equiv \frac{eB}{\omega}$ and introducing a UV-cutoff the small time differences, we obtain an expression for $m_p$:
\begin{align}
m_p(\omega) = m\(1 + g^2 \log \frac{\Lambda}{\omega} - i g^2 \frac{\Lambda}{\omega} \)
\end{align}
which is enhanced at small frequencies, and furthermore, diverges at $\omega = 0$. A signature of this effect is seen in the unconventional power law dependence of the cyclotron frequency of the polaron $\omega_{c}$ on $\Delta B$, discussed in the main text, which we obtain by solving for $\omega$ using $\omega = \frac{eB}{m_p(\omega)}$:
\begin{align}
\omega = \omega_{c,0} - g^2 \omega \log \frac{\Lambda}{\omega} + ig^2 \Lambda
\end{align}
The real and imaginary solutions of $\omega$ give $\omega_{c}$ and $\Gamma_c$ respectively, and have been stated in the main text. 

\subsection{Dephasing of quantum oscillations}
\label{app:dos}
The previous section focused on the energy-damping of oscillatory MHA-polaron motion due to radiation of soft-collinear MHA phonons. The cyclotron line-width computed there is relevant for phase-insensitive transport and optics cyclotron resonance measurements. 
For quantum oscillations measurements (e.g. Shubnikov-deHaas (SdH) oscillations in resistivity or related oscillations in tunneling density of states), the phase-coherence of the electronic orbit is also important, and these phenomena are affected more strongly by MHA phonons.

To analyze this effect, we employ a semiclassical computation of the density of states for an MHA polaron in an excess magnetic field $\Delta B$ away from commensurate filling, by summing up the ``return" amplitudes for a particle to start at some position, and return to that position time $t$ later, and then Fourier transforming with respect to $t$.

For cyclotron motion of an electron coupled to MHA-phonons, the electron picks up a Berry phase equal to:
\begin{align}
e^{i\theta_B}=e^{i\alpha \int dt ~\dot{\v{r}}\cdot \v{u}(\v{r}(t))}
\end{align}
This phase has quantum fluctuations due to the quantum fluctuations of $\v{u}$, which give a suppression factor for the quantum-oscillation amplitude $\sim e^{-\frac{\alpha^2}{2}\<\[\int dt \dot{\v{r}}\cdot \v{u}(\v{r}(t)) \]^2\>_{u}}$, where the average is taken over the thermal ensemble of phonons (we will restrict our attention to zero temperature). 

Since the phonons are much slower than the electrons, we can approximate the phonon configuration during a cyclotron orbit as being static (fluctuations of phonon field are approximately quenched on the time scale of cyclotron motion). In that limit, we can replace $\int dt \dot{\v{r}}\cdot \v{u}(\v{r}(t))\rightarrow \Phi$ where  (by Stoke's theorem)  $\Phi$ is the flux of the effective ``magnetic field"
\begin{align}
b=\nabla\times \v{u}
\end{align}
through the cyclotron orbit. This approximation is valid for up to $n\sim \frac{v_F}{c_s}$ cyclotron orbits, on longer time scales than $\frac{2\pi v_F}{c_s\omega_c}$, then we would have to account for the dynamics of the phonons.

Let us begin by computing the $\<uu\>$ correlator in real-space. We have a Matsubara action: 
$\frac{\rho}{2}\((\d_\tau \v{u})^2+c_s^2(\nabla \v{u})^2\)$, in $2+1d$ Euclidean time. At zero temperature, the Green's function of the phonon will just be the $3d$ Coulomb potential in $2+1d$ spacetime:
\begin{align}
D^M(\tau,\v{r}) = \frac{1}{2\pi \rho c_s\sqrt{c_s^2\tau^2+r^2}} \underset{\tau\ll r}{\approx}\frac{1}{2\pi\rho c_s}\frac{1}{r}.
\end{align}

The $\<bb\>$ correlator can be obtained by taking spatial derivatives of this propagator, giving: $\<bb\> = -\nabla^2\<uu\>=\frac{1}{2\pi\rho c_s r^3}$. Integrating over the cyclotron orbit, we get a flux-flux correlator:
\begin{align}
\<\Phi\Phi\> = \frac{1}{2\pi \rho c_s}\int_{\v{r},\v{r'}\subset \text{cyc. orbit}}  \frac{1}{|\v{r}-\v{r'}|^3} \approx \frac{\mathcal{A}_c}{4\pi \rho c_s}2\pi \int_a^{R_c}\frac{d\delta r}{\delta r^2} \approx \frac{\mathcal{A}_c}{\rho c_s a}
\end{align}
where $a$ is a short-distance (lattice-scale) cutoff, and 
\begin{align}
\mathcal{A}_c = \frac{S_k}{(e \Delta B)^2}
\end{align}
is the real-space area of the cyclotron orbit, and $S_k$ is the momentum-space area of the Fermi-surface.

Inserting this expression into the semiclassical version of the density of states, computed as a sum over return probabilities for various number of cyclotron oscillations~\cite{mirlin1996quasiclassical}, gives:
\begin{align}
N(\e) &= \frac{-1}{\pi}\int_0^t dt\text{Im}G^R(\v{r}=\v{r'};t)e^{-i\e t} \approx 
\<\sum_n e^{-i \frac{2\pi \e}{\omega_c}n}e^{-i\alpha n\Phi(\v{u})}\>_u
=	\sum_n e^{-i \frac{2\pi \e}{\omega_c}n}e^{-i\frac12\alpha^2 n^2\<\Phi\Phi\>}
\nonumber\\&=
\sum_n \sqrt{\frac{2\pi\rho c_sa}{\alpha^2\mathcal{A}_c}}\exp\[-\frac12\(\frac{\pi^2\rho c_s a}{\omega_c^2 \mathcal{A}_c\alpha^2}\)\(\e-n\omega_c\)^2\]
\end{align}
where in the last step, we have used the Poisson summation formula.

This result predicts a comb of Gaussian peaks centered at integer multiples of the cyclotron frequency with peak-amplitude: 
\begin{align}
A_c\approx \sqrt{\frac{\pi\rho c_sa}{\alpha^2\mathcal{A}_c}} \propto \Delta B
\end{align}
and width:
\begin{align}
\Gamma\sim\frac{\omega_c^2 \mathcal{A}_c\alpha^2}{\pi^2\rho c_s a}\propto \(\Delta B\)^{2g^2}.
\end{align}

By contrast, for elastic impurity scattering, the return probability for n-orbits is $e^{-n^2(\omega_c\tau)}$, and the quantum oscillation amplitude would scale like $\sqrt{\frac{1}{1/(\omega_c\tau)}}\sim \sqrt{\Delta B}$. (Note that the amplitude suppression of multiple scattering mechanisms combine in parallel not in series, so that the linear-$\Delta B$ dependence of the MHA-phonon dephasing would dominate at small $\Delta B$ over impurity contributions). Hence, the above linear-$\Delta B$ scaling provides a signature of the unconventional MHA polaron dynamics that is qualitatively distinct from conventional contributions from impurities.

We also remark that essentially identical expressions apply for analog systems of electrons coupled to emergent gauge-like fields, such as the composite fermion liquid of the half-filled Landau level, spinon Fermi surfaces in gapless $U(1)$ spin liquids, and zero-wavevector quantum critical points in metals.
\end{document}